\documentstyle[eqsecnum,aps,preprint,epsf]{revtex}
  
\begin{document}
 \thispagestyle{empty}
  \title{ Relativistic Structure of the Deuteron:\\  1. Electro-disintegration and y-Scaling}
      
 \author{ C. Ciofi degli Atti, D. Faralli,  A.Yu.  Umnikov and L.P. Kaptari
\thanks{On leave from Bogoliubov Lab. Theor. Phys., JINR, Dubna, Russia.}}
 \address{  Department of Physics, University of Perugia, and 
 INFN, Sezione di Perugia,
 via A. Pascoli, Perugia, I-06100, Italy.} 

 \date{\today}
 
 \maketitle

 
 \begin{abstract}
 Realistic solutions of the spinor-spinor Bethe-Salpeter equation
 for the deuteron with realistic interaction kernel
 including  the exchange of $\pi$, $\sigma$, $\omega$,
 $\rho$, $\eta$ and $\delta$ mesons, are used to
 systematically investigate relativistic effects
 in inclusive quasi-elastic electron-deuteron
 scattering within the relativistic impulse approximation.
 Relativistic $y$-scaling is considered by 
 generalising the non relativistic scaling function to
 the relativistic case, and it is shown that $y$-scaling
 does occur in the usual relativistic scaling variable
 resulting from the energy conservation in the instant
 form of dynamics. The present approach of $y$-scaling
 is fully covariant, with the deuteron
 being described by eight components, viz.
 the  $^3S_1^{++} $, $^3S_1^{--} $, 
      $^3D_1^{++} $, $^3D_1^{--} $, 
      $^3P_1^{+-} $  $^3P_1^{-+} $,
      $^1P_1^{+-} $, $^1P_1^{-+} $  waves.
      It is demonstrated that if the negative relative energy states 
 $^1P_1 $, $^3P_1$ are disregarded, the concept of covariant 
 momentum distributions ${\rm N}(p_0,{\bf p})$, with 
 $p_0=M_D/2-\sqrt{{\bf p}^2+m^2}$,
 can be introduced,
 and that calculations  of electro-disintegration cross section
 in terms of these distributions  agree within few percents with 
 the exact calculations which include the $^1P_1 $, $^3P_1$ states,  provided
the nucleon three momentum $|{\bf p}|\le 1 GeV/c$; in this momentum
range, the asymptotic relativistic scaling function is shown to coincide with the 
 longitudinal covariant momentum  distribution.
  
 \end{abstract}

\section{INTRODUCTION}
\label{sec:introduction}

The concept of scaling, which   plays an important role in the 
investigation of  hadronic structure,  can be introduced in the 
description of scattering processes  whenever the cross section 
 factorizes into a product of two different quantities, with  the first one
 reflecting the nature of
the scattering process, and therefore depending  
upon the relevant independent 
kinematical variables,  and the second 
one (the {\it scaling function})
reflecting the internal structure 
of the target, and  therefore depending upon 
 a new variable (the {\it scaling variable})
 which can be associated to the dynamics of the
 constituents of the target.

The best   example of scaling ($x$-scaling) is provided by inclusive 
  Deep Inelastic Scattering (DIS)
of leptons off nucleons~\cite{Bjork}: 
in the Bjorken limit ($\nu\to\infty,\, Q^2\to\infty$,
$x_{Bj}=Q^2/2m\nu={\rm const}$, 
with  $\nu$ and $Q^2$ being, respectively, the energy and 
the four-momentum transfers, and $m$  the nucleon mass), the 
quantity  $F_{2}(x_{Bj},Q^2) \equiv  \nu W_2^N(x_{Bj},Q^2) $
 where 
$ W_2^N(x_{Bj},Q^2) $ represents 
the deviation of the inclusive cross
 section from scattering off a point-like nucleon,
 becomes $Q^2$-independent, i.e. scales in the variable $x_{Bj}$
 (the Bjorken scaling variable) which can be associated to the
 momentum fraction of 
  the quark  inside the hadron. Inclusive
 scattering of leptons off a nucleus $A$
  in the Quasi-Elastic (QE) region
 ($\nu \le Q^2/2m $, i.e.  $x_{Bj}>1$) has been theoretically 
 shown~\cite{west,PaceSalme,ci,Donelly}
 to exhibit  another kind  of scaling, the so called
y-scaling, which can be summarized
as follows\footnote{
For  exhaustive reviews of $y$-scaling see refs.
\cite{west}, \cite{frankfu},\cite{ci} and \cite{Donelly}.}: 
at sufficiently high values of the three-momentum
transfer  ${\bf  q}$,
 the  quantity 
 $|{\bf  q}| W_{1(2)}^A (\nu, {\bf q}^2)/ W_{1(2)}^N (\nu, {\bf q}^2) $,
where the nuclear  structure function $W_{1(2)}^A(\nu, {\bf q}^2)$
 represents the deviation of the cross section
 from scattering off a point-like nucleus,
 scales to a function of the  variable  $y$, according to 
 $|{\bf q}| W_{1(2)} (\nu,{\bf q}^2)/ W_{1(2)}^N(\nu, {\bf q}^2) \to  F(y)$ 
 where, in the case of the
deuteron (but not for complex nuclei\cite{ci}), 
$F(y)$ represents the nucleon  longitudinal momentum distribution $f(y)$
\footnote{We will consider, from now on, negative values of $y$ for which the effects of non-nucleonic degrees of freedom are kinematically suppressed}
:
\begin{equation}
f(y) =
\int\limits_{0}^\infty\,n(p_\parallel,{\bf p}_\perp)\,d{\bf p}_\perp 
=2\pi \int\limits_{|y|}^\infty\,n({\bf p})\,|{\bf p}| \,d|{\bf p}|,\quad p_\parallel \equiv\ |y|.
\label{west} 
\end{equation}
\noindent
Experimental data, due to the effect of the final state interaction  (FSI),
exhibit only a qualitative scaling behaviour, and 
a quantitative  analysis\cite{ciofideut} of the deuteron data 
~\cite{Bosted}
  taking into  account (FSI), 
 allowed one to obtain the nucleon
momentum distribution in the deuteron using eq.  (\ref{west}).
It should however be stressed that 
the approach of ref.~\cite{ciofideut}, 
 leading to eq. (\ref{west}),
is based on a fully relativistic instant-form treatment of kinematics, but
on the usual non relativistic (Schr\"odinger) treatment of the deuteron,
and therefore lacks of a consistent covariant 
treatment of the process \footnote{
Note that FSI have been  treated in ref.~\cite{ciofideut} within a full Schr\"odinger equation approach, i.e. using ground and continuum eigenfunctions of the same Hamiltonian. Such an approach is a correct one at the kinematics of the data of Ref.~\cite{Bosted}  which, at negative values of $y$, correspond to a very small (less than the pion threshold) relative energy of the  neutron-proton pair in the continuum; it should   always be  kept in mind, however,  that at high values of $Q^2$, such that the nucleon-nucleon (NN) cross section becomes strongly absorptive,
the Schro\"odinger approach is  inadequate 
(see e.g. \cite{bianco})}.
 Recently \cite{polyzou}, 
 a relativistic covariant model, based on the  light-front dynamics
 and light cone kinematics  \cite{keister}, has been
 adopted  to analyze  inclusive 
 QE scattering off the deuteron, treating 
 the latter
  as a system of two spinless particles interacting via
 a simple scalar interaction.   
 Within such a model, the deuteron wave function has only one component
 (the $S$-wave),  
 the square of which defines the model momentum distribution.
 However, it is well known that in the realistic case of two interacting spinor
 particles, the deuteron state is determined by at least four 
 components if one nucleon is on mass shell \cite{gross}, five components 
 within the spinor light cone formalism \cite{karmanov},
 or even eight
 components within the covariant  
Bethe-Salpeter (BS) approach \cite{tjond,tjonn}. 

The necessity of 
 more than two components  ($S$ and $D$-waves)
 in the description of the deuteron
in  a covariant  approach, 
 follows from an accurate account of the contribution 
 of the
 negative relative energy states in the deuteron,  
the so-called $P$-waves.
  Frequently, in the calculation of observables 
  within covariant formalisms with
  realistic interactions, the final results are rather cumbersome
  (see, for instance, refs. \cite{gross,keisterTj,quad}) 
 and a separation, in a compact form,
 of an analog of the 
 momentum distribution of the deuteron becomes difficult,
 if not, sometimes, impossible. The inclusive quasi elastic cross section and
 the concept of  $y$-scaling have not been so far considered within 
 covariant approaches with realistic interaction, and an investigation of the
 possibility to  define  relativistic scaling functions and momentum distributions
 is still lacking.

In this paper we focus  on a detailed study of   quasi elastic
$eD$ scattering,
 and the possibility to  analyze it  in terms
of scaling functions and momentum distributions,
 using the numerical BS solution recently 
obtained with a realistic one-boson exchange (OBE)
interaction \cite{uk,upar,ukkk}. 

Through this paper, as in \cite{polyzou},
 relativistic effects will be
investigated within the Impulse Approximation (IA), which means
that the final state interaction of the $np$ pair in  
the continuum will be disregarded, though
it has been shown \cite{ciofideut} that FSI 
lead to sizeable scaling violation  effects at low values of $|{\bf q}|$.
 FSI effects will be  analyzed
in a subsequent paper \cite{elsewhere}, here we are only interested 
in the investigation of $y$-scaling within a fully covariant treatment
of inclusive $eD$-scattering in IA, within a description of the deuteron
in terms of realistic solutions of the BS equation.

Our paper is organized as follows:
in Section~\ref{sec:Basic}  
the derivation, within the BS formalism,
 of the two basic quantities which are necessary to define
 relativistic $y$-scaling, i.e. 
the cross section for elastic electron scattering from
a moving and off-mass-shell nucleon, and the cross section 
for inclusive quasi-elastic scattering from  the deuteron, are presented;
in Section \ref{sec:relativ} the relativistic scaling function is defined, 
its non relativistic reduction is illustrated, 
and the results of numerical calculations are presented;
 a relativistic momentum
distribution appropriate to the BS approach is 
defined in Section \ref{sec:relmom},
where the relationship between the relativistic
momentum distribution and the scaling function is
illustrated; 
the Summary and Conclusions
are presented in Section \ref{sec:concl};
 some relevant details concerning the construction of 
 the Mandelstam vertex
for the operator of $eD$ scattering and
for the computation 
 of  matrix elements 
 within the Bethe-Salpeter
formalism are given
in Appendices~ \ref{sec:operator} and \ref{sec:details}, 
 respectively.

\section{The electron-nucleon and electron -Deuteron  cross sections
within the Bethe-Salpeter formalism}
\label{sec:Basic}

\subsection{General formulae for the cross section}

In this Section the cross sections for the
  electron-hadron scattering
within the covariant BS formalism will be derived.
In particular, the elastic scattering
from
a moving and off-mass-shell nucleon, and  the inclusive quasi-elastic
scattering off the deuteron at rest, will be considered. Both processes
will be denoted $A(e,e')X$, where $A$ stands for the target hadron
and $X$ for the final hadronic states.
 The 4-momenta of the initial and final
electrons in the laboratory system are $k=({\cal E},{\bf  k})$ 
and $k'=({\cal E'},{\bf  k'})$, respectively;
 the four momentum transfer is  $q=k-k'=(\nu, {\bf  q})$, and 
the orientation of
the coordinate system is defined by  ${\bf  q}=(0,0,q_z)$.
 At high energies
 the electron mass can be disregarded, so that 
\begin{eqnarray}
k^2 = (k')^2 \simeq 0,\quad kk'= -kq = \frac{-q^2}{2} = \frac{Q^2}{2}, \label{kin10}\\
Q^2 \equiv - q^2 = 4{\cal E E}'\sin^2{\frac{\theta}{2}}.\label{kin11}
\end{eqnarray}
where $\theta$ is the scattering angle. The following relations will  be useful
in what follows:
\begin{eqnarray}
{\cal E} = \frac{\nu}{2}\left( 1+ \frac{
                       \sqrt{\sin^2{\frac{\theta}{2}}+\frac{Q^2}{\nu^2}}
                                        }{\sin{\frac{\theta}{2}}}\right),\quad
                                      {\cal E}' = {\cal E} -\nu \label{kin20}\\[1mm]
\cos{\theta_k} = \frac{1}{\sqrt{1+\frac{Q^2}{\nu^2}}}\left( 1+ \frac{Q^2}{2\nu {\cal E}}\right),   
\label{kin21}    \\
\left | {\bf  q}\right| = \left | q_z\right| =\sqrt{Q^2+\nu^2}, \label{kin22}                          
\end{eqnarray}
where $\theta_k$ is the polar angle of the 
initial electron.

Hereafter  the electron-nucleon vertex in the 
      on-mass-shell form, viz:
\begin{eqnarray}
\Gamma^{eN}_{\mu}(Q^2) = \gamma_{\mu} F_1(Q^2) + 
i \frac{\sigma_{\mu\alpha}q^{\alpha}}{2m}\kappa F_2(Q^2),
\label{vertex}
\end{eqnarray}
will be used, where $F_{1,2}$ are the 
electromagnetic form factors of the nucleon, and $\kappa$ its 
anomalous magnetic moment.
It  is well known that the choice (\ref
{vertex}) for $\Gamma^{eN}_{\mu}(Q^2)$ violates gauge invariance; this is a relevant  point which will be briefly discussed in  Section \ref{sec:relativ}.
In the one-photon exchange approximation
the general formula for the invariant
cross section for the process $A(e,e')X$ 
has the following form:
\begin{eqnarray}
d\sigma = \frac{1}{4kp_A}\left|  {\cal M}_{e+A \to e' + X }\right |^2
(2\pi)^4\delta^{(4)}\left( k+p_A -k'- p_{X} \right) \frac{d{\bf  k'}}{(2\pi)^3 2{\cal E}'}  
d\tau_X,
\label{cs00}
\end{eqnarray}
where $p_A$ is the initial 
target momentum, $p_{X}$ the total momentum of the final
hadron $(X)$ state, $d\tau_X$ the phase factor corresponding to the 
final hadron state $X$, and
${\cal M}_{e+A \to e' + X }$   
the invariant matrix element describing the
process. 
For the elastic electron nucleon ($eN$) (Fig.\ref{fig1}a)
 and electron deuteron  $(eD)$ scattering in the impulse approximation
 (Fig.\ref{fig1}b), 
 we have, respectively:
\begin{eqnarray}
\left.
\begin{array}{l}
p_A = p_1, \quad  p_{X} = p_1' = p_1+q,\\[2mm]
d\tau_X = \displaystyle\frac{d {\bf  p}_1'}{(2\pi)^3 2 E_1'}
\end{array}
 \right \} 
\quad {\mbox {for $eN$ scattering}}.
\label{eNf}\\[3mm]
\left.
\begin{array}{l}
p_A = P_D, \quad  p_{X} = p_1' + p_2 = P_D+q,\\[2mm]
d\tau_X = \displaystyle\frac{d {\bf  p}_1'}{\mbox{$(2\pi)^3 2 E_1'$}}
 \displaystyle\frac{d{\bf p}_2}{(2\pi)^3 2 E_2}, 
\end{array}
 \right \} 
\quad {\mbox{for $eD$ scattering}}.
\label{eDf}
\end{eqnarray}
Using the identity:
\begin{eqnarray}
 \frac{\mbox{$d$}{\bf  p}_1'}{\mbox{$(2\pi)^3 2 E_1'$}}       =
\delta\left((p_1')^2-m^2 \right)
\frac{\mbox{$d^4 p_1'$}}{\mbox{$(2\pi)^3 $}}, 
\label{eNf1}
\end{eqnarray}
the elastic $eN$ and the inclusive $eD$ cross sections 
are obtained by  integrating
 over the $d^4p_1'$:
\begin{eqnarray}
\frac{d\sigma}{d{\cal E'}d\Omega_{k'}} = \frac{1}{4kp_A}
\int\, \left| 
 {\cal M}_{e+A \to e' + X }\right |^2
(2\pi)\delta((p_1+q)^2-m^2) \frac{{\cal E}'}{2(2\pi)^3 } d\tau,
\label{cs01}
\end{eqnarray}
where $ d\tau$ is
the phase factor corresponding to the 
final hadron state $X$ without the hit nucleon, 
 $\Omega_{k'}$ is the
scattered electron solid angle and 
$p_1$  the initial nucleon momentum in $eN$
  ($p_1^2 = m^2$) and
 $eD$  ($p_1 = P_D-p_2$, $p_1^2 \neq m^2$) scattering.

The square of the invariant matrix element in eq.~(\ref{cs01}),
averaged over the spins of the colliding particles and summed over
the  spins
of the scattered particles, can be cast in the form:
\begin{eqnarray}
 \left|  {\cal M}_{e+A \to e' + X }\right |^2 (2\pi)\delta((p_1+q)^2-m^2) 
= \frac{ e^4}{Q^4} L^{\mu\nu}(k,q) W_{\mu\nu}^A(p_A,q),
\label{ime00}
\end{eqnarray}
 where the leptonic  tensor $L^{\mu\nu}$ has the familiar form:
\begin{eqnarray}
  L^{\mu\nu}(k,q) = 2 \left( 
  2k_\mu k_\nu -(k_\mu q_\nu +k_\nu q_\mu) + g_{\mu\nu}\frac{q^2}{2}
  \right).
\label{ime01}
\end{eqnarray}
As for the hadronic tensor $W^N_{\mu\nu}$, appearing in 
  elastic $eN$ scattering, one has
 
\begin{eqnarray}
 W^N_{\mu\nu}(q^2,p_1\cdot q)& = \frac{1}{2} {\sf Tr} \left \{
 (\hat p_1 +m)\Gamma_\mu^{eN}(Q^2)  (\hat p_1+\hat q +m) 
\Gamma_\nu^{eN}(Q^2) 
 \right \}(2\pi)\delta\!\left((p_1+q)^2-m^2\right).
\label{ime02}
 \end{eqnarray}
 By contracting (\ref{ime02}) with the leptonic tensor,
 eq. (\ref{ime01}), one obtains, 
  in the nucleon
rest system ( $ p_1=(m,{\bf 0})$),   the well known 
Rosenbluth cross section. 
In the case of a complex system, e.g., the deuteron , 
only the general expression  for   
the hadronic tensor can be unambiguously defined
 in terms of the two independent structure
functions, $W_{1,2}(q^2,p_D\cdot q)$,
whose explicit  form, however, relies on particular 
theoretical models; 
the model adopted in this paper
is described in the next subsection.

\subsection{The hadronic tensor}

Our strategy in computing the hadronic tensor
for a composite system  is the following one:
a {\em  nucleonic} tensor operator $\hat O^N_{\mu\nu}$
will be defined, whose expectation value
between relativistic hadronic states $\,|\,A\,\rangle$
generates  the corresponding hadronic tensor,
according to:
\begin{equation}
W_{\mu\nu}^{A} = \langle\, A\, | \hat O^N_{\mu\nu}\,|\,A\,\rangle.
\label{ime03}
\end{equation}
The general requirements for the 
operator $\hat O^N_{\mu\nu}$ are as follows:

\noindent
i) it should lead to eq. (\ref{ime02}) when sandwiched 
between free nucleon states ($\,|\,A\,\rangle = \,|\,N\,\rangle$);

\noindent
ii) when sandwiched 
between deuteron  states ($\,|\,A\,\rangle = \,|\,D\,\rangle$) it
should incorporate the effects from the Fermi motion and
the off mass shellness of the hit nucleon;

The operator $\hat O_{\mu\nu}$, 
 eq. (\ref{ime03}), due to the
 choice  of the vertex
(\ref{vertex}), has the following form:
\begin{eqnarray}
&&
\hat O^N_{\mu\nu}(p_1,q)= (2\pi)\delta\!\left((p_1+q)^2-m^2\right) 
\times\nonumber\\
&& 
\left\{ F_1^2(Q^2)\hat O_{\mu\nu}^{(1)}(p_1,q)+
\frac{\kappa}{2m} F_1(Q^2) F_2(Q^2)\hat O_{\mu\nu}^{(12)}(p_1,q)+
\frac{\kappa^2}{4m^2} F_2^2(Q^2)\hat O_{\mu\nu}^{(2)}(p_1,q)
\right \},
\label{ime030}
\end{eqnarray}
where
\begin{eqnarray}
&&\hat O_{\mu\nu}^{(1)}(p_1,q) = 2\left [
-g_{\mu\nu}(\hat p_1+\hat q-m)+\gamma_\mu p_{1\nu}+\gamma_\nu p_{1\mu}
\right ] ,
\label{ime031}\\[1mm]
&&\hat O_{\mu\nu}^{(12)}(p_1,q) =  4\left [
g_{\mu\nu}(-m\hat q + (q^2+qp_1))\right ],
\label{ime032}\\[1mm]
&&\hat O_{\mu\nu}^{(2)}(p_1,q) =  2\left [
g_{\mu\nu}(mq^2+\hat p_1 q^2 -\hat q(q^2+2qp_1))-
q^2(\gamma_\mu p_{1\nu}+\gamma_\nu p_{1\mu})
\right ],
\label{ime033} 
\end{eqnarray}
and all  terms proportional
to $q_{\mu(\nu)}$ have been omitted in view of the gauge invariance
of the leptonic tensor (\ref{ime01}),  $q_{\mu(\nu)}\,L^{\mu\nu}=0$
In eqs. (\ref{ime031})-(\ref{ime033}) and 
in  the rest of the paper
the short-hand notation $\hat p$ will be used  
 for the  scalar product of a four-vector $p$ with $\gamma$-matrices,
 i.e., $\hat p\equiv\gamma_\mu p^\mu$.  
In actual calculations we first contract $\hat O^N_{\mu\nu}$ with
the  leptonic tensor $L^{\mu\nu}$,
 and the resulting operator
is then sandwiched between target ground states. 
The result of the contraction is 
(see  Appendix~\ref{sec:operator}):
  
\begin{eqnarray}
&& \hat O(p_1,q,k) = L^{\mu\nu}\, \hat O_{\mu\nu}=
 \left\{ \hat O_{stat} +\delta \hat O_{mot} + \delta \hat O_{off}
\right \}(2\pi)\delta\!\left((p_1+q)^2-m^2\right),\label{ime07} 
\end{eqnarray}
where 
\begin{eqnarray}
&&\hat O_{stat} =2\left[ q^2m-4m\nu{\cal E} +
 4m {\cal E}^2 \right]\,A(Q^2)+\frac{q^4}{m}\,B(Q^2) ,
\label{ime066} \\
&&\delta \hat O_{mot} =
4\left[ m\nu {\cal E} -\hat q(kp_1)- {\cal E}(2m{\cal E} - m\nu) +\hat k(2kp_1-qp_1)\right]
\, A(Q^2),
\label{ime067} \\
&&
\delta \hat O_{off}\equiv \label{ime062}
\\[1mm]&&
 -2q^2\left( \hat q +\frac{q^2}{2m} \right) F_1^2(Q^2)+
 \frac{2\kappa q^2}{m} 
 \left( -m\hat q  +  qp_1 \right) F_1(Q^2)F_2(Q^2)-
 \frac{\kappa^2 q^2}{2m^2}
 \left[ \hat q  (q^2+2qp_1) \right] F_2^2(Q^2),
\nonumber
\end{eqnarray}
and 
\begin{eqnarray}
A(Q^2) &\equiv& \left( F_1^2(Q^2)- \frac{\kappa^2q^2}{4m^2}
 F_2^2(Q^2)\right ),
\label{scA}\\
B(Q^2) &\equiv& \left( \phantom{\frac{q^2}{m^2}}
 \!\!\!\!\!\!\!\!\! F_1(Q^2)+
\kappa F_2(Q^2)\right )^2.
\label{csB}
 \end{eqnarray}

Let us discuss the meaning of the  terms (\ref{ime066})-(\ref{ime062}):
$\hat O_{stat}$  represents the contribution from
the interaction of the lepton with the  nucleon at rest, and
in case of $eN$ scattering   its 
average over  nucleon spinors
yields  the Rosenbluth cross
section; $\delta\hat O_{mot}$ 
originates  from the motion of the nucleon;
 $\delta \hat O_{off}$  takes into account the 
off mass shellness  ($p_1^2\neq m^2$) of a bound nucleon,
 and therefore only apperas in the $eD$-cross section. 
  
As already pointed out, the $eN $  and $eD$ cross sections are obtained
by sandwiching the relativistic 
operator $ \hat O(p_1,q,k)$  between relativistic nucleon
or deuteron states; this means that, within 
  the present approach, the hadronic tensor has 
  exactly the same form 
  for a free or for a bound nucleon.
The only assumption we make,
 is that the electromagnetic vertices
 for free  and bound  nucleons are  of the same form, with
 all nuclear
   effects  provided by the  state vectors. 
 In the usual, non covariant
 PWIA, based on the use of non relativistic
 wave functions, the cross section off a nucleus
 $A$ is obtained by relating the nuclear
 hadronic tensor $W_{\mu\nu}^A$ to the nucleon
 hadronic tensor $W_{\mu\nu}^N$ by a convolution
 formula, with resulting ambiguities as 
 far as the extrapolation of the $eN$ cross section
 for a bound off-mass shell nucleon is concerned
 (see e.g. \cite{forest,jaus}).

As previously mentioned, gauge invariance is broken for  the deuteron current corresponding to the  choice 
Eq. (\ref{ime030}). Several phenomenological prescriptions have been
  suggested to restore it \cite{forest,jaus}. It should be pointed out that our paper is mainly aimed at   theoretically comparing relativistic and non relativistic deuteron momentum distributions and $y$-scaling functions,  without presenting any comparison with experimental data which would, of course, require a serious consideration of
gauge invariance violation effects.

 The procedure used in this paper can, 
 in principle, be adopted  
for the description of $eD$ scattering
 within  non relativistic  Schr\"odinger
 picture. However, in this case,
 consistency would require 
  a non relativistic reduction of the $\gamma NN$
  vertex.  A systematic study of  deuteron
 electro-disintegration within the non relativistic
 approach  taking into account 
 FSI,  relativistic
 corrections and  meson exchange currents,
 can be found in \cite{arenhovel} and
 references therein quoted.
\subsection{The elastic electron nucleon cross section}

The elastic $eN$ cross section resulting by sandwiching
$\hat O_{stat}$ and $\delta\hat O_{mot}$ between free nucleon states
 reads as follows: 
\begin{eqnarray}
&&
\frac{d\sigma}{d\Omega_{k'}} = \frac{{\cal E} E_{\vec p_1}}{(p_1k)}\,
f_{recoil}\, \widetilde{\sigma^{eN}},\label{moving}
\end{eqnarray}
where:

\noindent
i)  $f_{recoil}$ is the recoil term
\begin{equation}
f_{recoil}^{-1}=1+\frac{2{\cal E}\sin^2\theta/2-\nu}{E_{\vec p_1+\vec q}}-
\frac{p_{1z}}{|\vec q|}\,
\frac{\nu-2{\cal E}\sin^2\theta/2}{E_{\vec p_1+\vec q}}.  \label{recoil}
\end{equation}

\noindent
ii)
$\displaystyle\frac{{\cal E} E_{\vec p_1}}{(p_1k)}$ comes from the 
 redefinition of the
incident flux for a moving nucleon;
\noindent 
 iii)   $\widetilde{\sigma^{eN}}$  is  the ``reduced'' electron-nucleon 
 cross section, i.e. the  cross section
without the flux factor and the recoil contribution, namely

\begin{eqnarray}
\widetilde{\sigma^{eN}}\,&=&\,
 \sigma_{Mott}\,\frac{m^2}{E_{\vec p_1}E_{\vec p_1 +\vec q}}\times\nonumber\\[1mm]&&
\left\{
A(Q^2)+\frac{Q^2}{2m^2}\tan^2 \frac{\theta}{2}B(Q^2)+
\frac{A(Q^2)}{m^2{\cal EE}' \cos^2 \theta/2 }
\left[ (p_1k)^2-(p_1k)(p_1q)-m^2{\cal EE}'\right]\right\},
\label{cs021}
\end{eqnarray}
where 
\begin{eqnarray}
\sigma_{Mott} \equiv 
\frac{\alpha^2  \cos^2{\frac{\theta}{2}}}{4{\cal E}^2 \sin^4{\frac{\theta}{2}}}
\label{cs022}
\end{eqnarray}
is the Mott cross section.
In the above equation    $E_{\vec p_1}=\sqrt{m^2 + {\bf p}_1^2}$ ,
 $E_{\vec p_1+\vec q}=\sqrt{m^2 + ({\bf p}_1+{\bf q})^2}$.
 It can  easily be  seen from (\ref{moving})-(\ref{cs021})
that for a  nucleon  at rest ($p_1=(m,{\bf 0})$) eq.(\ref{moving}) coincides with
the Rosenbluth cross section.

 \subsection{The inclusive electron deuteron cross section within the BS formalism}
 
The {\it relativistic  impulse approximation}  for the 
 inclusive $eD$-cross section
 is obtained by  averaging  the operator $\hat O$,
given by eq. (\ref{ime07}), with the Bethe-Salpeter amplitudes (see 
 Fig.~\ref{fig2}). The result is

\begin{eqnarray}&&
\left (\frac{d\sigma}{d {\cal E}'d\Omega_{k'}}\right )^{BS}_{eD} =
  \sigma_{Mott} \,
  \int\limits_{|y|}^{p_{max}} 
       \frac{|{\bf  p}|^2d|{\bf  p}|}{(2\pi)^4}
        \frac{1 }{16{\cal E  E}' \cos^2 \frac \theta 2}
\int\limits_{0}^{2\pi}   
 \frac{d\varphi_p}{|{\bf  p}|\cdot|{\bf  q}|}
  \frac{1}{2wM_D}\frac{1}{(M_D(M_D-2w))^2}
    \nonumber \\[1mm]
&& \left[ \frac{1}{3}\sum_M 
{\sf Tr} \left \{ \bar G_M(p,P_D) {(\hat p_1 + m) }
\left( \hat O_{stat}+\delta \hat O_{mot}+\delta\hat O_{off}\right)
{(\hat p_1 + m) }
  G_M(p,P_D) (\hat p_2 + m) \right \}
  \phantom{\int\limits_{|y|}^{p_{max}}}
\!\!\!\!\!\!\!\!\!\! \right],
\label{cs050}
\end{eqnarray}
where $p_1 = P_D-p_2$,
$G_M$ is the Bethe-Salpeter $D\to NN$ vertex, 
 $M$ is the deuteron  total angular
momentum
projection,  $p$ is the relative momentum of the
nucleons, i.e. $p_{1,2} = P_D/2\pm p$.
 The deuteron-nucleon vertex  $G_M$ is the
truncated Green's function which is  related to the conjugated Bethe-Salpeter
amplitude  $\Psi_M$ by 
\begin{eqnarray}
\Psi_M(p,P_D) =  \frac{(\hat p_1 + m) }{\left( p_1^2-m^2\right )}
  G_M(p,P_D) \frac{(\hat p_2 + m)}{\left( p_2^2-m^2\right )}.
\label{bsa01}
\end{eqnarray}

The limits of integration in (\ref{cs050}), $|{\bf p}|_{min}\equiv |y|$ and
      $ p_{max}$ 
are obtained from  energy conservation provided by 
the $\delta$- function
\begin{eqnarray}
 \delta\!\left( (P_D-p_2+q)^2 -m^2\right ) 
&=& \frac{1}{2|{\bf  p}|\cdot|{\bf  q}|}\delta\! \left(\cos{\theta_p} -
\frac{M_D(M_D+2\nu)+q^2-2w(M_D+\nu)}{2|{\bf  p}|\cdot|{\bf  q}|}
 \right ),
\label{ime095}
\end{eqnarray}
which determines the value of $\cos\theta_p$ 
from the constraint:
\begin{eqnarray}
-1\leq \cos\theta_p \leq 1 .
\label{ime092}
\end{eqnarray}
Solving the inequalities (\ref{ime092}) with $\cos\theta_p$ 
defined by the argument of the  $\delta$-function (\ref{ime095}), we obtain the
same result as in \cite{ciofideut}, i.e.
\begin{eqnarray}
&&|{\bf p}|_{min} =  \frac{1}{2}\left |
\left \{ (M_D+\nu)\sqrt{1-\frac{4m^2}{s}} -|{\bf  q}|\right \}
\right |
\equiv \left | y \right|
\label{ime093} \\
&&|{\bf p}|_{max} = \frac 12 \left \{ (M_D+\nu)\sqrt{1-\frac{4m^2}{s}} 
+|{\bf  q}|\right \} \equiv p_{max},
\label{ime094}
\end{eqnarray}
where $s$ denotes the Mandelstam variable for the $\gamma^* D$ vertex
 \begin{equation}
s=(P_D+q)^2 =M_D(M_D+2\nu)-Q^2.
\label{sman} 
\end{equation}
It can be seen from (\ref{ime094}) that $ p_{max}$ sharply 
increases with $|{\bf q}|$;  such a circumstance, as we
shall see, has relevant consequences for the occurrence of $y$-scaling.

In the calculation of the trace 
appearing  in eq. (\ref{cs050}),
 two different operators have to be considered, namely 
  the scalar operator,
$\hat {\sf 1}$,  coming from  $\hat O_{stat} $( eq. (\ref{ime066})), and the
vector operator $\gamma_\mu$,  contained in $\delta \hat O_{mot}$ and
$\delta \hat O_{off}$, (eqs. (\ref{ime067}) and (\ref{ime062})). 

We  have calculated   the BS average of these operators, 
$\langle \hat {\sf 1}\rangle_{pole}$ 
and  $\langle \gamma_\mu \rangle_{pole}$,
 where the subscript  "pole" means that
  the trace in eq. (\ref{cs050}) has been
 obtained   using eq. (\ref{bsa01})
and  by evaluating the integral  over $p_{20}$  in the 
pole corresponding to the second particle on mass shell
(for details, see Appendix~\ref{sec:details}),  i.e.,
\begin{eqnarray}
p_{20} = w =\sqrt{{\bf  p}^2 + m^2}, \quad  p_0 = \frac{M_D}{2} - w, \quad p_{10} =
M_D - w.
\label{ime09}
\end{eqnarray}

The $eD$-cross section 
can now  be rewritten in terms of the nucleon pole contributions to
the vector and the scalar parts, namely 
\begin{eqnarray}
\left ( \frac{d\sigma}{d{{\cal E}'}d\Omega_{k'}} \right )^{BS}_{eD}
 &=&
 \sigma_{Mott} (2\pi)\,
 \int\limits_{|y|}^{p_{max}}        \frac{|{\bf  p}|^2d|{\bf  p}|}{(2\pi)^4}
 \frac{1 }{8{\cal E  E}'\cos^2\frac \theta 2}
              \int\limits_{0}^{2\pi}    \frac{d\varphi_p}{2|{\bf  p}|\cdot|{\bf  q}|}
\left( f_{stat}+\delta f_{mot}+\delta f_{off}\right),
\label{cs060}
\end{eqnarray}
where
\begin{eqnarray}
&& f_{stat}  =\left [
 2m\left[ q^2+4 {\cal E E}' \right]\,A(Q^2) +\frac{q^4}{m} \,B(Q^2)
\right ]
 \langle \hat{\sf 1} \rangle^{BS}_{pole}  ({\bf  p}),
\label{f1}\\
&& \delta f_{mot}= 
4
\phantom{\frac{q^2}{m^2}}\!\!\!\!\!\!\!\
\left[ 
 -2m  {\cal E E}'\, 
\langle \hat{\sf 1} \rangle^{BS}_{pole}  ({\bf  p}) 
 +\left( - q^{\mu}(kp_1) +  k^{\mu}(2kp_1-qp_1)\right )
 \langle \gamma_\mu \rangle^{BS}_{pole}  ({\bf  p})
  \right]\,A(Q^2),
\label{f2}\\
&& \delta f_{off}= 
 -2q^2\left[  q^{\mu}\langle \gamma_\mu \rangle^{BS}_{pole}  ({\bf  p})
  +\frac{q^2}{2m}\langle \hat{\sf 1} \rangle^{BS}_{pole}  ({\bf  p})
   \right] F_1^2(Q^2)\nonumber  \\
 &&\quad  \quad \quad \quad  \quad \quad+
 \frac{2\kappa q^2}{m} 
 \left[ -m  q^\mu\langle \gamma_\mu \rangle^{BS}_{pole}  ({\bf  p})
   +  qp_1 \langle \hat{\sf 1} \rangle^{BS}_{pole}  ({\bf  p})
   \right]F_1(Q^2)F_2(Q^2)\nonumber  \\
 &&\quad  \quad \quad \quad  \quad \quad -
 \frac{\kappa^2 q^2}{2m^2}
 \left[  q^\mu \langle \gamma_\mu \rangle^{BS}_{pole}  ({\bf  p})
   (q^2+2qp_1) \right] F_2^2(Q^2).
\label{f3}
\end{eqnarray}

Let us compare equation  (\ref{cs060}), which   represents the BS
$eD$ inclusive cross section in which the matrix elements of
both the $eN$ and $D\to NN$ vertices are treated covariantly,
with the usual non covariant 
PWIA  cross section, where the $eN$ vertex is
treated covariantly  within the
instant-form of dynamics, and the vertex $D\to NN$ is
 treated non covariantly
within the Schr\"odinger approach, 
viz.
 
\begin{equation}
\left ( \frac{d\sigma}{d{{\cal E}'}d\Omega_{k'}} \right )^{PWIA}_{eD}
=
 (2\pi)\,   \int\limits_{|y|}^{p_{max}}    |{\bf  p}|  \, d|{\bf  p}|
 n_D({\bf  p})  \frac{E_{\vec p+\vec q} }{|{\bf q}|}
 \left (\overline\sigma_{ep}+\overline\sigma_{en}\right ),
 \label{pwia}
 \end{equation}
 where $ \overline\sigma_{ep(n)}$ is the relativistic electron-nucleon cross section
 for a free moving nucleon 
 (see, e.g., ref. \cite{forest} ) and 
the non relativistic momentum distribution 
$n_D({\bf p})$ is normalized as follows:
\begin{eqnarray}
 \int\, d {\bf  p}\, n_D({\bf  p}) =1.
\end{eqnarray}

The most relevant difference  between eqs. (\ref{cs060}) and (\ref{pwia}) 
arises from the non covariant treatment 
 of the  $D\to NN$ vertex which,
in eq. (\ref{pwia}), 
is  entirely  determined by a 
single quantity,
the nucleon momentum distribution $n_D({\bf p})$, 
whereas in the covariant BS
cross section, eq.  (\ref{cs060}), it 
depends upon $\langle \hat {\sf 1}\rangle^{BS}_{pole}$
 and  $\langle \gamma_\mu \rangle^{BS}_{pole}$.
 In order to exhibit the quantitative difference between
 the two cross sections let us compare
 $\langle \hat {\sf 1}\rangle^{BS}_{pole}$
  and  $\langle \gamma_\mu \rangle^{BS}_{pole}$ with
  their non relativistic limits ($ |{\bf p}|^2 /m^2 << 1$)
  which are given by
  (for details  see refs.\cite{uk,nashphysrev}):
\begin{eqnarray}
\langle \gamma_\mu\rangle^{BS}_{pole} ({\bf p})
\to (2\pi)^3(p_\mu/E_{{\bf p}}) \,n_D({\bf  p}),
\quad
\langle {\sf 1} \rangle^{BS}_{pole} ({\bf  p})
\to (2\pi)^3 (m/E_{{\bf p}})\,n_D({\bf  p}),
\label{gpole}
\end{eqnarray}
where $ p_\mu =(E_{{\bf p}},{\bf  p})$ and $n_D({\bf  p})$ is the deuteron 
momentum distribution.
    
  In Figs.\ref{fig3} and \ref{fig4}  
 $\langle \hat {\sf 1}\rangle^{BS}_{pole}$
  and  $\langle \gamma_\mu \rangle^{BS}_{pole}$ 
  are compared with 
  their non relativistic limit obtained with $n_D({\bf p}) $ corresponding 
 to various  realistic interactions. 
Using eqs. (\ref{gpole})  the non relativistic limit of the BS cross section
 can be obtained straightforwardly, viz.
 
\begin{equation}
\left ( \frac{d\sigma}{d{{\cal E}'}d\Omega_{k'}} \right )^{NR}_{eD}
=
 (2\pi)\,   \int\limits_{|y|}^{p_{max}}    |{\bf  p}|  \, d|{\bf  p}|
 n_D({\bf  p})  \frac{E_{\vec p+\vec q} }{|{\bf q}|}
 \left (\widetilde\sigma_{ep}+\widetilde\sigma_{en}\right ),
\label{nonrel}
\end{equation}
where  
$\widetilde\sigma_{eN}=\widetilde\sigma_{eN}(|{\bf p}|, |{\bf q}|, \nu)$
 is given  by eq. (\ref{cs021}), without
the  off mass shell contribution
(eq. (\ref{f3})), which  can easily be  shown to vanish
 in the non relativistic limit.

Thus, we have demonstrated   that  the non relativistic limit
  of the BS inclusive cross section 
 (\ref{cs060}),  obtained by taking
 the non relativistic limit of 
 $\langle \hat {\sf 1}\rangle^{BS}_{pole}$
 and  $\langle \gamma_\mu \rangle^{BS}_{pole}$,
 has exactly the same structure as
 the instant form result of ref. \cite{ci},
apart from some minor differences between the relativistic 
$eN$ cross sections  $\widetilde{\sigma}_{eN}$ and  $\overline{\sigma}_{eN}$,
 which are irrelevant for the present paper, and  which will be
 discussed elsewhere \cite{elsewhere}.

In closing this Section, the following remarks are in order:

\noindent
i)  the BS covariant inclusive
$eD$ cross section does not factorize into a product  
of an electron nucleon cross section and a deuteron structure function.
In that respect the covariant results 
  differ from the usual non covariant PWIA;
 
\noindent
ii) within the BS formalism the interacting nucleon
is consistently treated as an off mass shell particle. 
Consequently, the matrix element of the $\gamma NN$ vertex
is half off shell for the $eD$ scattering. As a result, additional
off mass shell effects, represented by eq. (\ref{f3}),
 arise due to covariance of  the approach.

\section{The relativistic scaling function}
\label{sec:relativ}
\subsection{Non relativistic and relativistic 
scaling functions}

In the non relativistic case, 
the concept of $y$-scaling can be introduced when the value of 
 $|{\bf q}|$ becomes large enough so has to 
 make $p_{max}\sim \infty$ and the dependence of
 $\overline\sigma_{eN}$  ( or $\widetilde\sigma_{eN}$)
 upon $|{\bf p}|$ very weak. In such a case eq. (\ref{nonrel})  
  can be cast in the following form:
 
 \begin{equation}
\left ( \frac{d\sigma}{d{{\cal E}'}d\Omega_{k'}} \right )^{NR}_{eD}
=
 \left (s_{ep}+s_{en}\right )
 \frac{E_{y+|\vec q|} }{|{\bf q}|}\,
 (2\pi)\, 
   \int\limits_{|y|}^{\infty}|{\bf  p}|  \, d|{\bf  p}|
 n_D({\bf  p}),
 \label{nonrelappr}
 \end{equation}
where $s_{eN}$ and $E_{y +|\vec q| }$ represent 
$\widetilde\sigma_{eN}$ and 
$E_{\vec p +\vec q }$ ,
calculated   at $|{\bf p}|=|{\bf p}|_{min}=|y|$ and
 are taken out
of the integral. Such an approximation has been carefully 
investigated in ref. \cite{ci} and found to be 
valid within few percents, provided $Q^2 >0.5 GeV^2/c^2$.
 It is clear therefore, that  at large values of $|{\bf q}|$
  the following quantity
 (the non relativistic scaling function) 
\begin{eqnarray}
&&
F^{NR}(|{\bf q}|,y) \equiv \frac{|{\bf q}|}{E_{y+|\vec q |}}\cdot
\left(\frac{d\sigma}{d {\cal E}'d\Omega_{k'}}\right)^{NR}_{eD}/
\left(s_{ep} +s_{en}\right )
\label{scfunnon}
\end{eqnarray}
 will be directly related to the longitudinal momentum distribution~\cite{ci}

\begin{equation}
 F^{NR}(|{\bf q}|,y)\,
\longrightarrow \,f(y)
=2\pi\int\limits_{|y|}^{\infty} |{\bf p}| d |{\bf p}| n_D(|{\bf p}|),
\label{add1}
\end{equation}
whose first derivative will provide    the
 non relativistic momentum
distribution.
As already pointed out, the condition
for the occurrence of non relativistic $y$-scaling is that eq. (\ref{nonrel})
could be cast in the form (\ref{nonrelappr}), which means that:
i) $ Q^2 > 0.5 GeV^2/c^2 $, in order to make the replacement
$\widetilde \sigma_{eN}\to \, s_{eN}$ and 
$E_{\vec p+\vec q}\to\,E_{y+|\vec q|}$ possible, and
ii) $p_{max}=(|{\bf q}|-|y|)\gg  |y|$ (cf. (\ref{ime093}) and
(\ref{ime093})) in order to saturate the integral of the momentum
distribution, 
$\int\limits_{|y|}^{p_{max}} |{\bf p}| d|{\bf p}| n_D(|{\bf p}|)\to
 \int\limits_{|y|}^\infty |{\bf p}| d|{\bf p}| n_D(|{\bf p}|)$.
 Condition ii) obviously implies that the larger the value of
 $|y|$, the larger the value of $|{\bf q}|$ at which
 scaling will occur. The satisfaction of
 the inequalities $|{\bf q}|\gg 2|y|,\, x_{Bj} > 1 $
 leads, for any well-behaved $ n_D(|{\bf p}|)$, to the following
 conditions for the occurrence of non relativistic $y$-scaling:
 \begin{equation}
 2m/3\,\leq \nu\, < |{\bf q}|,\quad|{\bf q}|\geq 2m.
 \label{conditions}
 \end{equation}
  Note, that
 the above conditions are very different from the conditions 
 for Bjorken scaling $\nu \simeq |{\bf q}|$.
 Let us now discuss relativistic scaling. 
To keep contact with non relativistic
scaling, let us define the following relativistic  
  scaling function:
\begin{eqnarray}
&&
F^{BS}(|{\bf q}|,y) \equiv \frac{|{\bf q}|}{E_{y+|\vec q |}}\cdot
\left(\frac{d\sigma}{d {\cal E}'d\Omega_{k'}}\right)^{BS}_{eD}/
\left( s_{ep} +s_{en}\right ),
\label{scfun}
\end{eqnarray}
with
$\left(\displaystyle\frac{d\sigma}{d {\cal E}'d\Omega_{k'}}\right)^{BS}_{eD}$
 given by eq. (\ref{cs060}). In the rest of the paper the following questions will
be dadresse:

\noindent 
i) does (and at which values of $|{\bf q}| $) eq. (\ref{scfun}) 
 scales in $y$ ?
 
 \noindent 
 ii) if  scaling does occur, 
 can a relationship be established between
 the asymptotic scaling function
 and the momentum  distribution?

It is clear, by looking at eq. (\ref{cs060}),
that relativistic scaling, as the non relativistic one,
is entirely governed by the value of $p_{max}$ (eq. (\ref{ime094})),
in that one expects that, starting from a certain
high value of $|{\bf q}|$, $p_{max}$ becomes 
large enough as to saturate the $|{\bf p}|$ dependence
of $\langle \hat {\sf 1}\rangle_{pole}(|{\bf p}|) $ 
and  $\langle \gamma_\mu \rangle_{pole}(|{\bf p}|)$.
It should be pointed out that if
scaling of  (\ref{scfun}) is observed, this would imply that
a factorisation of (\ref{cs060}) similar to the one
occurring in the non relativistic case (i.e. eq. (\ref{scfunnon}))
occurs in relativistic scaling as well. Such a factorisation, due 
  the complex structure of  eq.  (\ref{cs060}) is, a priori, 
  not obvious. As for the second question, it is
by no means a trivial one, for in the BS case
  even the concept of  momentum distributions is not well defined.
Nevertheless, we will see that the  concept of relativistic
momentum distribution can be introduced, and that a relationship
of such a momentum distribution
with the asymptotic scaling function, can be established.

\subsection{ Numerical calculations of the relativistic cross section and scaling function}
\label{ssec:numeric}
  
In this section the  results of numerical calculations 
of the relativistic scaling function
 $F^{BS}(|{\bf q}|,y)$,  eq. (\ref{scfun}),
  will be  presented .
 In our calculations  the numerical
solution \cite{uk,ukkk} of the spinor-spinor BS equation
containing a  realistic one-boson-exchange interaction kernel, which includes
the set of $\pi$, $\sigma$, $\omega$, $\rho$, $\eta$ and $\delta$
 exchanged mesons, is used.
The meson parameters (masses, coupling constants and cut-off parameters)
have been taken  to be the same as in ref.~\cite{tjond,tjonn}, except for 
the coupling constant of the scalar $\sigma$-meson, which has been adjusted 
to provide a numerical solution of the homogeneous BS equation. 
Recently, the solution became available 
in the form of analytical parametrizations
 obtained by fitting the numerical solution
to the BS equation, using the least-squares procedure~\cite{upar}.

The details of the numerical calculation of  various matrix elements
appearing in $eD$ electro-disintegration
are given in Appendix \ref{sec:details}. 

In Fig.\ref{fig5} the approach to scaling of the BS scaling function 
(\ref{scfun}) is shown for various values of $y$.
  It can be  seen that scaling is approached very rapidly due to 
the sharp increase of   $ p_{max}$  in eq. (\ref{cs060})
 with $| {\bf q}|$ (cf. Fig.\ref{fig6}). 
It can also be seen that the value of $|{\bf q}|$, at which scaling
is reached,  sharply increases with the value of $y$,
 going from $|{\bf q}|\sim 1 GeV/c$  ( $\nu\sim 0.3 GeV$ ),
 at $y=-0.2 GeV/c$, to $|{\bf q}|\sim 2 GeV/c$
   ($\nu\sim 0.8 GeV$) at $y=-0.8 GeV/c$.
These values match very well the condition for  non relativistic 
$y$-scaling (\ref{conditions});
this apparently surprising result will be explained later on.
Let us now briefly discuss the role played by inelastic channels
on the scaling function. 
It can be shown \cite{elsewhere} that for all values
of $y$ presented in Fig.\ref{fig5},  the values of $\nu$ 
and $|{\bf q}|$ in the region of the approach to scaling
are below the pion production threshold, which can be
reached only at higher values of $|{\bf q}|$, provided
$Q^2\leq 5\,\, GeV^2/c^2$.
  Therefore, inelastic excitations of the nucleon are
 kinematically forbidden in a wide range of $|{\bf q}|$
 for a given value of $y$. 
   The asymptotic scaling function is shown vs. $y$   in Fig.\ref{fig7}, whereas the 
contribution of the off mass shell
terms (\ref{f3}) is presented in Fig.\ref{fig8}. 
 By comparing the two figures, it can be seen   
 that the off mass shell corrections are negligibly 
small in the whole range of $|y|$, and in what follows
they will be  disregarded. Figure \ref{fig9}
illustrates the role of  the ``moving'' components,
eq. (\ref{f2}), calculated  at different values of $|{\bf q}|=3, 10$ and 
$18\,\, GeV/c$; as expected,  these corrections increase
with $|y|$  and are practically $|{\bf q}|$-independent. 
The various contributions to the total  scaling function
are presented  in  Fig.\ref{fig10},
whereas, in   Fig.\ref{fig11} the
asymptotic relativistic BS scaling function is compared with
the  non relativistic one, calculated with various 
realistic interactions. It can be seen that for $|y|\,>\,0.4\,\, GeV/c$,
the differences between the BS and the non relativistic scaling 
functions are very large, except for  the RSC interaction.

 The scaling behaviour of the relativistic
 scaling function (\ref{scfun}) shown in Fig.\ref{fig5} would imply
 the factorisation of the BS inclusive cross section
 (\ref{cs060}) into the free $eN$ cross section
 and some kind of deuteron structure function. 
 Due to the complexity of eqs. (\ref{f1})-(\ref{f3}),
 neither the origin of such a factorization,
 nor the nature of the deuteron structure function
 are clear at the moment; they  will  however be
  discussed and clarified in the next Section.
   
   \section{Relativistic momentum distribution}
\label{sec:relmom}
Since in  the covariant BS formalism the deuteron amplitude
 does not have a probabilistic interpretation, the concept 
 of momentum distribution is ambiguous.
 Nevertheless, we will now rearrange the matrix elements
 of $\hat{\sf 1}$ and $\gamma_\mu$ in such a way that, 
 under certain conditions, they could be interpreted in
 terms of a relativistic  momentum distribution.
  
 Let us return to the main quantity, eq. (\ref{cs050}), and 
 try to analyze it analytically
in more details.
To this end, it is convenient in the 
decomposition of the BS 
amplitude,  to shift  from the Dirac basis, 
used in ref. \cite{uk}, to a
 basis of spin-angular matrices \cite{tjond,cubis},
 i.e.
an outer product of two spinors, representing 
the solutions of the
free Dirac equation with positive and negative energies.
 This basis, which is frequently used,  is labeled by the
 relative momentum $\vec p$, the helicities
 $\lambda_i$ and the energy spin $\rho_i$ of the
  particles~\cite{tjond},
and is  sometimes  called the 
 $(J,\lambda_1,\lambda_2,\rho_1,\rho_2)$ representation. 
 The spectroscopic
notations are used for
  the partial amplitudes
 viz.  $^{2S+1}L^{\rho_1,\rho_2}_J$, i.e.,
 \begin{equation}
 ^3S^{++}_1, \,^3S^{--}_1,\,
  ^3D^{++}_1,\, ^3D^{--}_1, \, ^1P^{+-}_1,
 \, ^1P^{-+}_1,\, ^3P^{+-}_1, \,^3P^{-+}_1.
   \label{spectr}
   \end{equation}
  The partial amplitudes in the basis (\ref{spectr})
 exhibit a more transparent physical meaning,  since they can 
 be compared with the deuteron states in the non relativistic limit.
 It is intuitively clear
 that the two nucleons in the deuteron are mainly in positive energy
 states with $L=0,2$,  so that one may
 expect  the probability of negative energy states
 with  $L=1$ in
 eq.~(\ref{spectr}) to be much smaller than
  the probability for the
 $^3S^{++}_1$ and $^3D^{++}_1$ 
 states.
 Moreover, it can be shown that the waves
 $^3S^{++}_1$ and $^3D^{++}_1$  directly correspond to the
 $S$ and $D$ waves in the deuteron, 
 with the waves with negative energy vanishing
 in the non relativistic limit.   
(The connection between the partial amplitudes defined in the Dirac
and the  $\rho$-spin representations can be found in ref. \cite{quad}).

Let us now investigate analytically the matrix elements
$\langle\,\gamma_\mu\,\rangle_{pole}^{BS}({\bf p})$
and  $\langle\,{\sf 1}\,\rangle_{pole}^{BS}({\bf p})$,
eqs. (\ref{f1})-(\ref{f3}). Their explicit expressions are:

\begin{eqnarray}
&& {\cal R}_{\hat O}({\bf p})\equiv \nonumber \\[1mm]
&&\left .\left (
-i \frac{1}{6M_D} \sum_M  \frac{1}{(p_1^2-m^2)^2} \frac{1}{(p_2^2-m^2)}
Tr \left[\bar G_M\,(\hat p_1+m)\hat O \,(\hat p_1+m)G_M\,(\hat p_2+m)\right ] 
\frac{d p_0}{2\pi} \right )\right|_{\rm pole},
\label{pol1}
\end{eqnarray}
where $\hat O $ stands for either $\gamma_\mu$ or
${\sf 1}$.
Now  instead of calculating the pole
 contributions in eq. (\ref{pol1}), 
we go back to the  original definition of these
averages,   
following  eqs.(\ref{cs050})-(\ref{bsa01}), namely
\begin{equation}
{\cal R}_{\hat O}({\bf p}) =
\left .\left (
\frac{(p_2^2-m^2)}{4M_D w} \, \frac 13 \sum_M
Tr \left[\bar \Psi_M\,\hat O \,\Psi_M\,(\hat p_2-m)\right ] 
 \right )\right |_{p_0 = M_D/2 -E_2},
\label{trace}\\[2mm]
\end{equation}
and   calculate directly
the trace (\ref{trace}) evaluated at $ p_0 = M_D/2 -E_2$.
  Here  it is worth emphasizing that 
in our  case,  when one particle (the second one in the present  notation) is
on mass shell,  only four partial amplitudes contribute to the 
process \cite{gross}, namely, only those partial amplitudes (\ref{spectr})
with the second $\rho$-spin index positive, i.e.
the $ ^3S^{++}_1,\,  ^3D^{++}_1,\, ^1P^{-+}_1$ and $ \,^3P^{-+}_1$ amplitudes.
In correspondence to these contributions,
 we introduce  the Bethe-Salpeter wave functions
 for each vertex, viz.

\begin{eqnarray}
\Psi_S(p_0,|{\bf p}|)=
 \frac{1}{\sqrt{2}\pi} 
\, \frac { G^{++}_S(p_0,|{\bf p|})/2\pi}{\sqrt{2M_D}(2E_2-M_D)};\quad
&&
\Psi_D(p_0,|{\bf p|})=
 \frac{1}{\sqrt{2}\pi}
\, \frac { G^{++}_D(p_0,|{\bf p}|)/2\pi}{\sqrt{2M_D}(2E_2-M_D)};
\label{big}\\[3mm]
\Psi_{P_1}(p_0,|{\bf p}|)=
 \frac{1}{\sqrt{2}\pi} 
\, \frac { G^{-+}_{^1P_1}(p_0,|{\bf p}|)/2\pi}{\sqrt{2M_D}M_D};\quad
&&
\Psi_{P_3}(p_0,|{\bf p}|)=
 \frac{1}{\sqrt{2}\pi} \, \frac { G^{-+}_{^3P_1}(p_0,|{\bf p}|)/2\pi}{\sqrt{2M_D}M_D},
\label{small}
\end{eqnarray}
where the normalization factors have been chosen so as to correspond to the
non relativistic normalization of the deuteron wave function:

\begin{eqnarray}
\int  d{\bf p} \left (
u^2 ({\bf p}) + w^2 ({\bf p})\right ) =1.
\label{normnr}
\end{eqnarray}

Then for ${\cal R}_{\hat O}$ we obtain:
\begin{eqnarray}
&&
{\cal R}_{\gamma_0} = (2\pi)^3\,\left (
\Psi_S^2(p_0,|{\bf p}|)+\Psi_D^2(p_0,|{\bf p}|)+
\Psi_{P_1}^2(p_0,|{\bf p}|)+ \Psi_{P_3}^2(p_0,|{\bf p}|) \right );
\label{gamma0}\\[2mm]
&&
{\cal R}_{\vec \gamma} = (2\pi)^3\,\frac{{\bf p}}{E_2}
 \left (
\Psi_S^2(p_0,|{\bf p}|)+\Psi_D^2(p_0,|{\bf p}|)-
\Psi_{P_1}^2(p_0,|{\bf p}|)- \Psi_{P_3}^2(p_0,|{\bf p}|) \right )
+\delta {\cal R}_{\vec\gamma} ;
\label{gammavec}\\[2mm]&&
{\cal R}_{\hat {\sf 1}} = (2\pi)^3\,\frac{{m}}{E_2}
 \left (
\Psi_S^2(p_0,|{\bf p}|)+\Psi_D^2(p_0,|{\bf p}|)-
\Psi_{P_1}^2(p_0,|{\bf p}|)- \Psi_{P_3}^2(p_0,|{\bf p}|) \right )
+\delta {\cal R}_{\hat {\sf 1}}\label{hatI},
\end{eqnarray}
where

\begin{eqnarray}
&&
\delta {\cal R}_{\vec\gamma} =  (2\pi)^3\,\frac{{\bf p}}{E_2}
\left \{
\frac{2\sqrt{3} m}{3|{\bf p}|}
\left [
\Psi_S(p_0,|{\bf p}|) \left(\Psi_{P_1}(p_0,|{\bf p}|) - \sqrt{2} \Psi_{P_3}(p_0,|{\bf p}|)
\right ) \right .\right.
\nonumber \\[2mm]
&&
\phantom{\delta {\cal R}_{\vec\gamma} =  (2\pi)^3\,\frac{{\bf p}}{E_2}
  \frac{2\sqrt{3} m}{3|{\bf p}|}         }
\left.\left.
 +\Psi_D(p_0,|{\bf p}|) \left( \sqrt{2}\Psi_{P_1}(p_0,|{\bf p}|) +\Psi_{P_3}(p_0,|{\bf p}|)
\right) \right ] \right \};
\label{veccorr}\\[2mm]
&&
\delta {\cal R}_{\hat {\sf 1}} = - (2\pi)^3\,\frac{{m}}{E_2}
\left \{
\frac{2\sqrt{3}|{\bf p}| }{3m}
\left [
\Psi_S(p_0,|{\bf p}|) \left(\Psi_{P_1}(p_0,|{\bf p}|) - \sqrt{2} \Psi_{P_3}(p_0,|{\bf p}|)
\right ) \right .\right.
\nonumber \\[2mm]
&&
\phantom{\delta {\cal R}_{\vec\gamma} =  (2\pi)^3\,\frac{{\bf p}}{E_2}
  \frac{2\sqrt{3} m}{3|{\bf p}|}         }
\left.\left.
 +\Psi_D(p_0,|{\bf p}|) \left( \sqrt{2}\Psi_{P_1}(p_0,|{\bf p}|) +\Psi_{P_3}(p_0,|{\bf p}|)
\right) \right ] \right \};
\label{scalcorr}
 \end{eqnarray}

As a conclusion, the inclusive $eD$ cross section will be given
by eq.  (\ref{cs060}) with eqs. (\ref{f1})-(\ref{f3}) for
$\langle \hat 1\rangle^{BS}$  and 
 $\langle \gamma_\mu \rangle^{BS}$  
 replaced by eqs.
 (\ref{gamma0})-(\ref{scalcorr}).

From what we have exhibited, 
it can be  seen that in the BS formalism, there is no  universal
 momentum distribution (cf eqs.  (\ref{gamma0}) - (\ref{hatI}))
  so that, in principle, 
 a factorized  cross section in 
the form  (\ref{nonrel}) does not hold. This is a consequence of the
covariance of the BS formalism, where   the 
small components (\ref{small}) 
with negative relative energies are taken into  account. 
 It has been shown\cite{quad}   that the contribution from 
  the waves with
positive relative energies are much larger than the one from
  $\Psi_P^2(p_0,|{\bf p}|)$, which 
    therefore, can be disregarded.  
However, the  corrections  (\ref{veccorr}) and
(\ref{scalcorr}),  resulting  from the interference
 between large  and small
waves contribute  both to the static and  to the moving  nucleon
contribution to  the $eD$ cross section and, ``a priori'',
 cannot be disregarded; accordingly they 
    will be taken
into account in our calculations. 
  Let us now introduce the following quantity, which will be
  called hereafter {\it the covariant relativistic 
  momentum distribution} :
  
\begin{equation}
{\rm N}^{BS}(p_0,{\bf p})= {\rm N}(p_0,{\bf p})+\delta {\rm N}(p_0,{\bf p}),
\label{distrBS}
\end{equation}
where
\begin{eqnarray}
&&
{\rm N}(p_0,{\bf p}) =\left (
\Psi_S^2(p_0,|{\bf p}|)+\Psi_D^2(p_0,|{\bf p}|) \right );
\label{reldisr}\\[1mm]
&&
\delta {\rm N}(p_0,{\bf p})=
\left \{
\frac{2\sqrt{3} }{3}
\left [ \phantom{\frac 11}
\Psi_S(p_0,|{\bf p}|) \left(\Psi_{P_1}(p_0,|{\bf p}|) - \sqrt{2} \Psi_{P_3}(p_0,|{\bf p}|)
\right ) + \right. \right. 
\nonumber\\[1mm]
&&
\hspace*{3cm}\left.\left.\phantom{\frac 12} 
\Psi_D(p_0,|{\bf p}|) \left( \sqrt{2}\Psi_{P_1}(p_0,|{\bf p}|) +\Psi_{P_3}(p_0,|{\bf p}|)
\right) \right ] \right \}.
\label{deltan}
\end{eqnarray}
Since the relative energy $p_0$ is fixed,
the  momentum distribution (\ref{reldisr}), which is 
defined only in terms of  the $S$ and $D$ components,
 resembles  
 the non relativistic distribution $n_D({\bf p})$,  
and therefore it is expected to 
provide  the main contribution to the $eD$ cross section. 
 
Thus the matrix elements ${\cal R}_{\hat O}$, eqs. (\ref{gamma0})-(\ref{hatI})
 can be written in 
the following way:
\begin{eqnarray}
&&
 {\cal R}_{\gamma_\mu} =  (2\pi)^3\,\frac{p_\mu}{E_2}\,
\cdot\left \{ 
\begin{array}{cc}
{\rm N}(p_0,{\bf p}), \quad\quad  & \quad\quad \mu=0; \\
{\rm N}(p_0,{\bf p}) + \displaystyle\frac{m}{|{\bf p}|} 
\delta {\rm N}(p_0,{\bf p}), &\quad \mu =(1,2,3)
\end{array}   \right.
\label{distr1}\\[2mm]
&&
 {\cal R}_{\hat {\sf 1}} = (2\pi)^3\frac{m}{E_2}\cdot\left\{
{\rm N}(p_0,{\bf p}) - \frac{|{\bf p}|}{m}\delta {\rm N}(p_0,{\bf p})\right\}.
\label{distr2}
\end{eqnarray}

If,
 for the time being,  the contribution of $\delta {\rm N}(p_0,{\bf p})$
is disregarded , it is possible to relate the BS inclusive $eD$
cross section to the elastic $eN$ cross section
for a  moving nucleon; as a matter of fact,  by 
inserting eqs.  (\ref{distr1})-(\ref{distr2}) 
into eqs. (\ref{f1})-(\ref{f3}),
one obtains:
 
\begin{equation}
\left ( \frac{d\sigma}{d{{\cal E}'}d\Omega_{k'}} \right )^{BS}_{eD}
=
 (2\pi)\,   \int\limits_{|y|}^{p_{max}}    |{\bf  p}|  \, d|{\bf  p}|
{\rm N}(p_0,{\bf p})    \frac{E_{\vec p+\vec q} }{|{\bf q}|}
 \left (\widetilde\sigma_{ep}+\widetilde\sigma_{en}\right ),
\label{secBS}
\end{equation}
At large values of  $|{\bf q}|$
eq.  (\ref{secBS}) becomes

\begin{eqnarray}
&&
\left(
\frac{d\sigma}{d{{\cal E}'}d\Omega_{k'}} \right)_{eD}^{BS}
\approx 
\left \{ s_{ep} + s_{en} \right \}
\frac{ E_{|{\vec q}|+y}}{|{\bf q}|}
(2\pi)
\int\limits^{\infty}_{|y|} |{\bf p}| d |{\bf p}|
{\rm N}(p_0,{\bf p}).
\label{approxscaling}
\end{eqnarray}
where, as before,  
  $\widetilde\sigma ^{eN}$ and
  $E_{\vec p+\vec q}$ have been evaluated at  
$|{\bf p}| =|{\bf p}_{min}| = |y|$.
 If  eq. (\ref{approxscaling}) 
 is placed in (\ref{scfun}), the BS
 asymptotic scaling
function is obtained, viz.
 
\begin{eqnarray}
&&
f^{BS}(y)\equiv \frac{|{\bf q}|}{E_{y+ |\vec q|}}
\left(
\frac{d\sigma}{d{\cal E}'d\Omega_{k'}} \right )^{BS}_{eD}\cdot
\left \{ 
s_{ep} +s_{en} \right \}^{-1} 
=(2\pi)\int\limits_{|y|}^\infty |{\bf p}| d |{\bf p}|
{\rm N}(p_0,{\bf p}).
\label{relf}
\end{eqnarray}
from which information on the covariant nucleon momentum distribution 
 ${\rm N}(p_0,{\bf p})$ could be obtained.
             
The asymptotic  scaling function, calculated by  eq. (\ref{relf}),
coincides  with the exact scaling function  (obtained using 
   eqs. (\ref{cs060})-(\ref{f3}) in eq. (\ref{scfun}))  in the whole range of
   $y$ (the largest  difference occurring at $y=-0.8$  and being less than 10\%),
    which means that the BS inclusive cross section
 could be safely replaced by its approximation (\ref{approxscaling}).
 In order to understand  such a result, in  Figs. \ref{fig12}- \ref{fig14}
  the separate BS waves, viz. the $S$ and $D$ waves (full lines of Figs. \ref{fig12} and \ref{fig13}), and the $P_1$ and $P_3$ waves (Fig.  \ref{fig14}), are presented.  It can be seen that the latter waves are smaller by order of magnitudes than the former ones, so that the quantity $\delta {\rm N}(p_0,{\bf p})$, due to the 
 interference terms 
(\ref{veccorr}) and (\ref{scalcorr}) generated by the  negative energy states, 
turns out to be negligible  up to   $|{\bf p}|\sim 1 GeV/c$.
 Thus we can conclude that  in the interval
    $0 <|{\bf p}|< 1 GeV/c $ 
the total distribution
 (\ref{distrBS}) can be safely approximated  by the diagonal 
 contribution  (\ref{reldisr}), as illustrated by Fig. \ref{fig15}, where the full  momentum distributions are shown.
 This is  the reason why the BS inclusive
  cross section factorizes in the same way as the
  non relativistic one, and the relativistic scaling function
  (\ref{scfun}) scales in $y$.
In Figs. \ref{fig12}- \ref{fig15} we have also compared the full BS results with the results from other approaches, viz. the non relativistic Schr\"odinger approach with different types of interactions (see also ref. \cite{quad}), as well as the relativistic approach based upon the Gross equations \cite {gross}.
The  BS and Gross  approaches differ both in the form of the relativistic equations as well as 
in the number of exchanged bosons considered in the kernel (six in the former approach and four in the latter one), but both reproduce equally well the  experimental NN phase shifts and the ground state properties of the deuteron, which is reflected in the very similar behaviour of the $S$ and $D$ waves  shown in Figs.\ref{fig12} and  \ref{fig13}; these results also show that the high momentum content ($|{\bf p}| \simeq  0.5 GeV/c$) generated by the relativistic equations is appreciably higher than the one provided by non relativistic wave functions.

\section{Summary and Conclusions}
\label{sec:concl}
In the present paper the inclusive quasi-elastic electron-deuteron
cross section has been analyzed within the 
relativistic Impulse Approximation using recent, realistic
solutions\cite{ukkk} of the spinor-spinor Bethe-Salpeter equation
for the deuteron, with the  interaction kernel
including the exchange of
 $\pi$, $\sigma$, $\omega$,
 $\rho$, $\eta$ and $\delta$ mesons. In our  approach,
  both the $\gamma^*N$ and the $D\to NN$ vertices
 are treated relativistically, with eight components for the deuteron
 amplitude, unlike 
 the usual, non relativistic approach\cite{ci}, in which the  
$\gamma^*N$ is described by a relativistic free electron-nucleon
cross section, and the $D\to NN$ vertex by the usual non relativistic
two-component  Schr\"odinger  wave function 
(we reiterate that when we  call the latter approach 
"non relativistic",  we refer  only to the
$D\to NN$ vertex) . The aim of our
paper was twofold, viz.: i) to investigate the relevance
of  relativistic effects, and ii) to understand whether the
concept of  $y$-scaling can be introduced in a relativistic 
description of inclusive  $eD$-scattering. The main results
 of our analysis can be summarized as follows:
 
1. The relativistic inclusive $eD$ cross section has been obtained in terms
     of the pole matrix elements  $\langle \hat {\sf 1}\rangle^{BS}_{pole}(|{\bf p}|) $ 
and  $\langle \gamma_\mu \rangle^{BS}_{pole}(|{\bf p}|)$,
 taking  into account the off-mass shellness of the nucleon
 and it has been found
 that unlike the non relativistic case, the BS cross section does not
 factorize into a product of the free electron-nucleon cross section and
 a structure factor depending upon  the deuteron momentum
 distribution.

2.  
It has been shown that the BS cross section can be written as a function 
of  the three momentum transfer ${\bf q}$ and a variable $y$, 
 which is exactly the same relativistic scaling variable
 used in the  non 
 relativistic approach and resulting from the relativistic instant-form energy
 momentum conservation. Thus,
 in full analogy with the non relativistic case, a relativistic scaling function
 $F^{BS}(|{\bf q}|, y)$ has been  defined as the  ratio
 of the BS $eD$ cross section to   the free $eN$ cross section
  (times a proper phase space factor), and 
  $y$-scaling of  $F^{BS}(|{\bf q}|, y)$
   has been demonstrated to occur,
  i.e. $F^{BS}(|{\bf q}|, y)\to f^{BS}( y)$, with 
  the conditions for relativistic $y$-scaling being  very similar to
 those of non relativistic scaling, 
 i.e.  $2m/3\,\leq \nu\, < |{\bf q}|,\quad|{\bf q}|\geq 2m$.
 
3.
It has been pointed out, that 
 whereas the mechanism of non relativistic scaling is easily understood
in terms of the rapid decay of the momentum distribution $n_D({\bf p})$,
which makes the non relativistic scaling function
$F^{NR}(|{\bf q}|,y) \sim
2\pi \int\limits_{|y|}^{|{\bf q}|-|y|}\,n({\bf p})\,|{\bf p}| \,d|{\bf p}|
$
to rapidly saturate with $|{\bf q}|$, i.e. to scale in $y$, a similar
explanation in the relativistic case is not, in principle, 
possible, since, as stated in point 1, the BS cross section does not factorize,
and, moreover,   the concept of
momentum distribution in the BS case is not uniquely defined.
Thus, in order to understand the mechanism of the observed 
 relativistic $y$-scaling, the role of the various components
of the BS amplitude was analyzed, and it has been found
that if  the extremely small diagonal contribution 
of the negative energy $P$-waves is omitted, it
is possible to define a covariant momentum
distribution of the form 
 ${\rm N}^{BS}(p_0,{\bf p})= {\rm N}(p_0,{\bf p})+\delta {\rm N}(p_0,{\bf p})$,
 where 
  $\delta {\rm N}(p_0,{\bf p})$, which originates from
 the interference between the  positive and negative waves, can be safely
 disregarded  provided $|{\bf p}| < 1 GeV/c$, so that,
 as a result, the BS cross section factorizes in the same way as
  does the  non relativistic cross section and  
  the relativistic scaling function becomes
$F^{BS}(|{\bf q}|,y) \sim
2\pi \int\limits_{|y|}^{|{\bf q}|-|y|}\,  {\rm N}(p_0,{\bf p})\,|{\bf p}| \,d|{\bf p}|
$;
such a result provides the explanation for the relativistic
$y$-scaling  and makes it possible  to obtain 
the BS covariant momentum distributions
 by a simple first order derivative of the asymptotic
BS scaling function
  $f^{BS}(y) \sim
2\pi \int\limits_{|y|}^{\infty}\,  {\rm N}(p_0,{\bf p})\,|{\bf p}| \,d|{\bf p}|
$. 

4.  The BS relativistic momentum distribution,  
${\rm N}(p_0,{\bf p})$, and the non relativistic one , $n_D({\bf p})$,
are practically the same 
up to $|{\bf p}| \sim 0.4\div 0.5 \,\,GeV/c$, where they start to differ by an amount which depends upon the two-body interaction producing $n_D({\bf p})$.

 To sum up, it can be concluded that,
 if  the effects from  negative energy $P$-states
 can be disregarded, which has been demonstrated to be
 the case when the nucleon momentum
 in the deuteron $|{\bf p}| \le 1  GeV/c$,   the concept of
 $y$-scaling can be introduced in the BS relativistic description
 of inclusive quasi-elastic $eD$ scattering, in the same way as 
 it is in  
 the conventional non relativistic  approach,
 i.e. by introducing a scaling function which,  in the scaling regime, 
 is nothing but the nucleon longitudinal 
 momentum distribution; moreover, both in the relativistic and non 
 relativistic cases,   scaling  is shown to occur in the
 same variable $y$, and at values of $\nu$ and $|{\bf q}|$
 such that quasi-elastic scattering is the 
dominant process.

\acknowledgments

 C.d.A. is very grateful to F. Gross for supplying us with the numerical
values of the wave functions shown in Figs.\ref{fig12} and \ref{fig13}.
L.P.K. and A.Yu. U. are indebted  to INFN,
  Sezione di Perugia,  for  warm hospitality and financial support. 
  This work has been partially supported by RFBR grant
No 96-15-96423.

\appendix
\section{The construction of the nucleon operator}
\label{sec:operator}

 The contraction of the operators 
 (\ref{ime030})-(\ref{ime033})
with the leptonic tensor (\ref{ime01}) is given by:
\begin{eqnarray}
&& \hat O(p_1,q,k)\equiv \hat O_{\mu\nu}(p_1,q)L^{\mu\nu}(k,q) 
 \label{ime0400}\\[2mm]
 &=& \left\{ F_1^2(Q^2)\hat O^{(1)}(p_1,q,k)+
\frac{\kappa}{2m} F_1(Q^2) F_2(Q^2)\hat O^{(12)}(p_1,q,k)+
\frac{\kappa^2}{4m^2} F_2^2(Q^2)\hat O^{(2)}(p_1,q,k)
\right \}\\
&&\quad \times(2\pi)\delta\!\left((p_1+q)^2-m^2\right),
\label{ime040}
\end{eqnarray}
where
\begin{eqnarray}
&&\hat O^{(1)}(p_1,q,k) = 2\left [
q^2m-\hat q(q^2+2kp_1)+\hat k(4kp_1-2qp_1)
\right ],
\label{ime041}\\[1mm]
&&\hat O^{(12)}(p_1,q,k) =  4q^2\left [
(-m\hat q  + (q^2+qp_1))
\right ],
\label{ime042}\\[1mm]
&&\hat O^{(2)}(p_1,q,k) = 2q^2 \left [
mq^2 -\hat q  (q^2+2qp_1-2kp_1)-2\hat k (2kp_1-qp_1)
\right ].
\label{ime043} 
\end{eqnarray}
The average of (\ref{ime043})  with the nucleon or the deuteron amplitudes,
times the factor $e^2/Q^4$, gives the
invariant matrix elements.

A graphical representation of  the operator $\hat O(p_1,q,k)$ is
presented in Fig.~\ref{app1}, where the 
crossed nucleon line corresponds to a nucleon 
on the mass-shell 
with the propagator  
$(2\pi)\delta\!\left  ((p_1+q)^2-m^2\right )(\hat p_1 + \hat q +m)$.

The nucleon operator  $\hat O(p_1,q,k)$ is the  central object in our discussion
about the connection between the nucleon and the deuteron cross sections. 
Let us first consider 
 the differential elastic cross section for the
 process $e+N\to e'+ N$. 
The invariant matrix element is defined by
\begin{eqnarray}
 \left|  {\cal M}_{e+N \to e' +N }\right |^2 (2\pi)\delta\! \left((p_1+q)^2-m^2\right) 
&=& \frac{ e^4}{Q^4}  \frac{1}{2}\sum_{s_1} \langle p_1, s_1 | \hat O(p_1,q,k) | p_1, s_1  \rangle
\label{ime50}\\
&=& \frac{ e^4}{Q^4}  \frac{1}{2}{\sf Tr} \left \{ (\hat p_1 +m)\hat O(p_1,q,k) \right \}.
\label{ime51}
\end{eqnarray}
Let us first obtain  the cross section in the rest frame of the nucleon
($p_1=(m,{\bf 0})$)
\begin{eqnarray}
   \frac{1}{2}{\sf Tr} \left \{ (\hat p_1 +m)\hat q \right \} =2m\nu, \quad
   \frac{1}{2}{\sf Tr} \left \{ (\hat p_1 +m)\hat k \right \}=2m{\cal E},\quad
   \frac{1}{2}{\sf Tr} \left \{ (\hat p_1 +m)       \right \}=2m,
\label{ime52}\\
kp_1 = m{\cal E}, \quad qp_1 = m\nu,\label{ime53}
\end{eqnarray}
so that 
 \begin{eqnarray}
&& \left|  
{\cal M}_{e+N \to e' +N }\right |^2  (2\pi) \delta\! \left((p_1+q)^2-m^2\right) =
\nonumber\\[1mm]
&& 
\frac{ e^4}{Q^4} 
 \left[4\left( m^2q^2-m\nu q^2-4m^2\nu{\cal E} +4 m^2{\cal E}^2\right)F_1^2(Q^2)
+  4\kappa q^4 F_1(Q^2) F_2(Q^2)+
 \phantom{\frac{\kappa^2}{4m^2}}
   \right.\nonumber \\[1mm]
&&  
\quad\quad\quad\quad\quad\quad \left.
{\kappa^2 q^2} \left( q^2 -4  {\cal E}^2 +4  \nu {\cal E} \right)F_2^2(Q^2)
\phantom{\frac{q^2}{m^2}} \!\!\!\!\!\!\!\!\!\right](2\pi)\delta\!\left(2m\nu-Q^2 \right)=
\label{ime54}\\[1mm]
&&\quad \quad
 \frac{ e^4}{Q^4} 16m^2{\cal E  E}' \left[
\left( \cos^2{\frac{\theta}{2}}-\frac{q^2}{2m^2}\sin^2{\frac{\theta}{2}} 
\right)F_1^2(Q^2)-\frac{\kappa q^2}{m^2}\sin^2{\frac{\theta}{2}} F_1(Q^2) F_2(Q^2)-
 \phantom{\frac{\kappa}{4m^2}}
   \right.\nonumber \\[1mm]
&&
\quad\quad\quad\quad\quad\quad\quad\left.
 \frac{\kappa^2q^2}{4m^2} \left(  1+\sin^2{\frac{\theta}{2}} \right)F_2^2(Q^2)
\right](2\pi)\delta\!\left(2m\nu-Q^2 \right)=
\label{ime55}\\
&& \frac{ e^4}{Q^4} 16m^2{\cal E  E}'
 \left[
\cos^2{\frac{\theta}{2}} \left( F_1^2(Q^2)- \frac{\kappa^2q^2}{4m^2} F_2^2(Q^2)\right )
-\frac{q^2}{2m^2}\sin^2{\frac{\theta}{2}} 
\left( \phantom{\frac{q^2}{m^2}} \!\!\!\!\!\!\!\!
F_1(Q^2)+\kappa F_2(Q^2)\right )^2
\right]\,\times \nonumber\\[1mm]
&&
(2\pi)\delta\!\left(2m\nu-Q^2 \right).
\label{ime56}
\end{eqnarray}

Inserting the last expression in  eq.~(\ref{cs01})
we get the Rosenbluth cross section:
\begin{eqnarray}
&&
\frac{d\sigma}{d{\cal E'}d\Omega_{k'}} = 
\frac{\alpha^2 m}{2{\cal E}^2 \sin^4{\frac{\theta}{2}}}\times\nonumber\\[1mm]
&&
\left[
\cos^2{\frac{\theta}{2}} \left( F_1^2(Q^2)- \frac{\kappa^2q^2}{4m^2} F_2^2(Q^2)\right )
-\frac{q^2}{2m^2}\sin^2{\frac{\theta}{2}} \left( \phantom{\frac{q^2}{m^2}} \!\!\!\!\!\!\!\!\!
F_1(Q^2)+\kappa F_2(Q^2)\right )^2
\right]\delta\!\left(2m\nu-Q^2 \right)=
\label{cs02a}\\[1mm]
&&
 \sigma_{Mott}
\left[
 \left( F_1^2(Q^2)- \frac{\kappa^2q^2}{4m^2} F_2^2(Q^2)\right )
-\frac{q^2}{2m^2}\tan^2{\frac{\theta}{2}} \left( \phantom{\frac{q^2}{m^2}} \!\!\!\!\!\!\!\!\!
F_1(Q^2)+\kappa F_2(Q^2)\right )^2
\right]\delta\!\left(\nu-\frac{Q^2}{2m} \right)
\label{cs021a}
\end{eqnarray}

Let us now consider the $eD$ cross section. To this end,
we will consider an arbitrary reference frame
where the four momentum of the nucleon
is $p_1=(p_{10},{\bf p}_1)$; moreover the
nucleon can be off mass shell ( $p_1^2\neq m^2$).
 
Our strategy now is to explicitly separate that  part of 
 the operator 
describing the free nucleon at rest, i.e. the part defining 
the cross section (\ref{cs021a}), from the remaining parts of the
operator due to the nucleon motion
and the off-mass-shell corrections.

 Rearranging the terms in 
eqs.~~(\ref{ime0400})-(\ref{ime043}) according  to their contributions, we
find
\begin{eqnarray}
&& \hat O(p_1,q,k) =  \left\{ \hat O_{on} + \delta \hat O_{off}
\right \}(2\pi)\delta\!\left((p_1+q)^2-m^2\right)\label{ime060} 
\end{eqnarray}
where $  \hat O_{on}$ is the sum of terms contributing 
to  the
invariant matrix element  for the free nucleon, and 
$\delta \hat O_{off}$ is the sum of terms 
providing the off-mass-shell corrections.
Since we can add to and subtract
from   both terms in eq.~(\ref{ime060}) pieces 
vanishing for a nucleon on the mass-shell, the separation
 on $\hat O_{on}$ and
 $\delta \hat O_{off}$ is not unique, but the physical results
do not depend on the way this is done. In our definition:
\begin{eqnarray}
&&
\hat O_{on}\equiv\frac{q^4}{m} \, B(Q^2) 
+ 2\left[ q^2m-2\hat q(kp_1) +2\hat k(2kp_1-qp_1)\right]\,A(Q^2),\label{ime061} \\[1mm]
&&
\delta \hat O_{off}\equiv \label{ime062app}\\[1mm]&&
 -2q^2\left( \hat q +\frac{q^2}{2m} \right) F_1^2(Q^2)+
 \frac{2\kappa q^2}{m} 
 \left( -m\hat q  +  qp_1 \right) F_1(Q^2)F_2(Q^2)-
 \frac{\kappa^2 q^2}{2m^2}
 \left[ \hat q  (q^2+2qp_1) \right] F_2^2(Q^2).
\nonumber
\end{eqnarray}

It can be seen that in eq.~(\ref{ime061})
$A(Q^2)$ and $B(Q^2)$ are weighted by two different functions, viz. a scalar, and  a mixture of scalar and vector terms, respectively. This is a result of the 
relativistic structure of the our formalism, in which, unlike the non relativistic case and the relativistic case for a nucleon at rest, the  vector charge density does not coincides with  the probability density.
 Because of the different structure of the weighting factors of $A(Q^2)$ and $B(Q^2)$  in $\hat O_{on}$, the average value of the latter calculated for the deuteron, will not factorize into a common term for $A(Q^2)$ and $B(Q^2)$. 
However, if the difference between
the vector  and the scalar charges is not too large,
one could be able to define a common term  for 
both $A(Q^2)$ and $B(Q^2)$, plus proper  correction terms. Therefore,
we redefine  $\hat O_{on}$ in the following way:
\begin{eqnarray}
 &&\hspace*{3cm} 
 \hat O_{on} \equiv \hat O_{stat} + \delta \hat O_{mot}, \label{ime065}\\
&&\hat O_{stat} =2\left[ q^2m-4m\nu{\cal E} + 4m {\cal E}^2 \right]\,A(Q^2)+\frac{q^4}{m}\,B(Q^2) ,
\label{ime066app} \\
&&\delta \hat O_{mot} =
4\left[ m\nu {\cal E} -\hat q(kp_1)- {\cal E}(2m{\cal E} - m\nu) +\hat k(2kp_1-qp_1)\right]
\, A(Q^2),
\label{ime067app} 
\end{eqnarray}
where $\hat O_{stat}$ is a 
scalar operator defining the invariant matrix element for the nucleon
at rest, whereas $\delta\hat O_{mot}$, which  is  defined by the combination of scalar  and
vector currents, gives a non-zero contribution
for the moving nucleon.

\section{The pole structure of the matrix elements within the  BS formalism}
\label{sec:details}

In computing the deuteron observables within
the BS formalism, i.e. the matrix elements of a given operator $\langle\, D| \hat O |D\,\rangle\equiv  \langle \hat O \rangle$, 
one makes use of the Mandelstam technique, which yields
\begin{eqnarray}
\langle \hat O \rangle
   &=&  \frac{i}{2M_D}\int \frac{d^4p}{(2\pi)^4} 
\frac{1}{(p_1^2-m^2+i\epsilon)^2(p_2^2-m^2+i\epsilon)} 
\nonumber \\[1mm]
&& \quad\quad
\frac{\mbox{$1$}}{\mbox{$3$}}\sum_M {\sf Tr}\left\{
\bar G_M(p_0,{\bf  p})
 (\hat p_1+m)  \hat O (\hat p_1+m)
 G_M(p_0,{\bf  p}) (\hat p_2+m)
\right \}.
\label{hato}
\end{eqnarray}
It can be seen that in eq. (\ref{hato}) there are poles and cuts in the integration variable 
 $p_0$.  However, the whole matrix element  
$\langle \hat O \rangle$ is real and finite. This allows one to perform
the Wick rotation in the complex plane of the relative energy $p_0$,
and to safely compute the integrals in eq. (\ref{hato}). Moreover, in 
this case, the BS vertex function $G_M(p_0,{\bf  p})$ depends upon
the imaginary part of $p_0$, which allows one to use directly the numerical
solutions  obtained in the rotated system \cite{ukkk,upar}.
In particular, when $\langle \hat O \rangle$ is defined at
fixed value of $p_0$ (which is just the case of  $eD$ 
processes investigated in this paper), the Wick rotation is no
longer relevant in the computation of matrix elements. In order to
establish a connection between the calculation of the matrix elements in
the form  (\ref{hato}) and the matrix elements (\ref{cs050}), we introduce the deuteron densities (see, also  ref.\cite{ukk})
\begin{eqnarray}
\langle \hat O \rangle^{BS}  ({\bf  p})
   &=&  \frac{i}{2M_D}\int \frac{dp_0}{(2\pi)} 
\frac{1}{(p_1^2-m^2+i\epsilon)^2(p_2^2-m^2+i\epsilon)} 
\nonumber \\[1mm]
&& \quad\quad
\frac{\mbox{$1$}}{\mbox{$3$}}\sum_M {\sf Tr}\left\{
\bar G_M(p_0,{\bf  p})
 (\hat p_1+m)  \hat O (\hat p_1+m)
 G_M(p_0,{\bf  p}) (\hat p_2+m)
\right \},
\label{densit}
\end{eqnarray}
where the integration contour is shown in Fig. \ref{app2}. There are two kinds
of singularities in eq. (\ref{densit}): i)when one of the particles
is on mass shell, $p_{10}=w$ or  $p_{20}=w$ labeled 
 ``1'' and ``2''  in Fig. \ref{app2}, 
 and ii)when both particles are deeply virtual, 
 $p_{10}=-w$ or  $p_{20}=-w$ ,labeled  ``3'' and ``4'',  respectively.
As previously mentioned,  in
eq. (\ref{densit}) the singularities are removed by   
performing the Wick rotation and by integrating along the imaginary
axis of $p_0$.  Instead of rotating the  contour, we close it in the
upper semi-plane, and integrate eq. (\ref{densit}) in the Minkowsky space.
Thus  there remain two singularities contributing to the full densities,
a  simple pole at  $p_{20}=w$, and a pole of  second order
  at $p_{10}=-w$. The former, which corresponds to the spectator on the mass-shell,
   gives the dominant contribution  to the full integral. It is exactly this
contribution which enters  
 our formulae in Section~\ref{sec:Basic}  (eq.  (\ref{cs050})), and
 in Section~\ref{sec:relativ} (eqs. (\ref{f1})-(\ref{f3})). 
  The approximation of the full matrix elements by the nucleon pole
  contribution is often used in the description of processes involving  the 
  deuteron~\cite{gross,shakin}. Accordingly, let us define the nucleon pole
  contribution to the full density:
\begin{eqnarray}
\langle \hat O \rangle^{BS}_{pole}  ({\bf  p})
   &=&  \frac{1}{2M_D} 
\frac{1}{2w M_D^2(M_D-2w)^2 }
\nonumber \\[1mm]
&& \quad\quad
\frac{\mbox{$1$}}{\mbox{$3$}}\sum_M {\sf Tr}\left\{
\bar G_M(p_0,{\bf  p})
 (\hat p_1+m)  \hat O (\hat p_1+m)
 G_M(p_0,{\bf  p}) (\hat p_2+m)
\right \}.
\label{Opole}
\end{eqnarray}
It can be seen that the matrix elements (\ref{cs050}) is the
pole part of (\ref{hato}). In so far as the pole contribution
(eq. (\ref{Opole}))  dominates  the full density (eq. (\ref{densit})),
the relevant quantities (eqs. (\ref{f1})-(\ref{f3})) can be calculated
by using the Wick rotation and by computing numerically the
integral (\ref{densit}) in the rotated system, using the numerical
solutions of the BS equation obtained in ref. \cite{ukkk}.
Fig.~\ref{app3} shows the charge $\langle \gamma_0\rangle$ and
scalar $\langle \sf 1 \rangle$ densities computed by eqs. (\ref{densit}) and  
(\ref{Opole}). It can be seen that up to $|{\bf p}|\sim 0.65 GeV/c$,
both methods provide the same results (above $0.65 GeV/c$ 
the pole contribution has been calculated  using  for the BS vertex function $G_M$,
 the analytical parameterization from ref.~\cite{upar}. More details on the behaviour of various full and pole 
densities can be found in ref. \cite{ukk}
    
\newpage

\begin{figure}     
\protect\caption{Diagrams corresponding
to  elastic $eN$ scattering (a),
and  inclusive $eD$ scattering in Impulse 
Approximation (b).}
\label{fig1}
\end{figure}

\begin{figure}     
\protect\caption{The  diagram corresponding to the
averaging of the operator $\hat O$,
eq. (\protect\ref{ime07} ), between deuteron states.
 The crossed line corresponds to
the nucleon on the mass-shell.}
\label{fig2}
\end{figure}

\epsfxsize 8cm
\begin{figure}     
\protect\caption{
The Bethe-Salpeter 
matrix elements $\langle\,{\sf 1}\,\rangle_{pole}^{BS}({\bf p})$
and  $\langle\,\gamma_0\,\rangle_{pole}^{BS}({\bf p})$  (solid lines)
compared 
with their non relativistic limits,
eq. (\protect\ref{gpole}), computed with the nucleon momentum
distribution $n_D({\bf p})$ corresponding to the 
 Bonn\protect\cite{bonn} (dashed), 
 Paris\protect\cite{paris} (dotted) 
 and Reid\protect\cite{Reid} (dash-dotted)
interactions.
}
\label{fig3}
\end{figure}

\epsfxsize 8cm
\begin{figure}     
\protect\caption{
The same as in Fig.\ref{fig3} but for the
vector density $\langle\,\gamma_3\,\rangle_{pole}^{BS}({\bf p})$}
corresponding to  $\cos \theta_{\widehat{\vec p\vec q}} = 1.$
\label{fig4}
\end{figure}

\begin{figure}  
\protect\caption{
The   scaling function 
$F^{BS}(|{\bf q}|,y)$, eq. (\protect\ref{scfun}) vs
$|{\bf q}|^2$  for various values of $y$. 
  For the sake of completeness, the value of the 
Bjorken variable $x_{Bj}= Q^2/2m\nu$ 
is also shown.}
\label{fig5}
\end{figure}

\begin{figure} 
\protect\caption{The  dependence of  the upper limit
of integration $|{\bf p}|_{max}$ in
eq. (\protect\ref{ime094}) upon $|{\bf q}|$, for fixed values of
$y$.}
\label{fig6}
\end{figure}

\begin{figure}  
\protect\caption{The  asymptotic scaling function
 $f^{BS}(y)$ computed
within the Bethe-Salpeter formalism (eq. (\protect\ref{scfun})
evaluated at $|{\bf q}|\to\infty$).}
\label{fig7}
\end{figure}

\begin{figure}  
\caption{Off-mass-shell contributions to 
$ f^{BS}(y)$:
the  three curves correspond (from the top)
 to the three terms of eq. (\protect\ref{f3}).}
\label{fig8}
\end{figure}

\begin{figure}   
\protect\caption{The ``moving corrections'' 
to $F^{BS}(|{\bf q}|,y)$, eq.  (\protect\ref{f2}),
 for $|{\bf q}| = 3, 10 $ and $18\,\, GeV/c$ respectively.
 The solid line is the asymptotic scaling function  $f^{BS}(y)$.}
\label{fig9}
\end{figure}

\begin{figure}    
\protect\caption{
 The various contributions to the
 asymptotic scaling function $f^{BS}(y)$.
 Dot-dased line:  the static part 
   (\protect\ref{f1});  dotted line: 
    the  "moving" corrections (\protect\ref{f2}).
         The solid line is the total scaling function 
         (eq.    (\protect\ref{scfun}) evaluated at $|{\bf q}|\to\infty$).
 }
\label{fig10}
\end{figure}

\begin{figure}   
\protect\caption{ 
 The BS asymptotic scaling function (full) compared 
 with the non relativistic scaling function (\protect\ref{scfunnon})
corresponding to the Reid (dashed), Paris (dotted) and
Bonn (dot-dashed) interactions. }
\label{fig11}
\end{figure}
 \begin{figure}   
\protect\caption{
The Bethe- Salpeter $S$-wave $|\, \Psi_S(p_0,|{\bf p}|)\, |$
(solid line), with 
$p_0=M_D/2- \sqrt{{\bf p}^2+m^2}$,
compared with the Gross $S$-wave  corresponding to the IIB solution
\protect\cite{gross} (dot-dashed) and with the  
 non relativistic deuteron  $S$-wave obtained from the
Paris (dotted) and Bonn (dashed) potentials. 
}
\label{fig12}
\end{figure}
\begin{figure}   
\caption{The same as in Fig. \protect\ref{fig12} for the
deuteron $D$-wave. 
}
\label{fig13}
\end{figure}

\begin{figure}   
\caption{The Bethe- Salpeter negative
relative energy waves  $\Psi_{P_{1,3}}(p_0,|{\bf p}|)$.
}
\label{fig14}
\end{figure}

\begin{figure}   
\caption{
The Bethe-Salpeter covariant 
 momentum distribution (eq.(\ref{distrBS})
 with $p_0=M_D/2- \sqrt{{\bf p}^2+m^2}$) (full)
 and the contribution from the negative relative energy
 states (\protect\ref{deltan}) (dot-dashed).
 The non relativistic momentum distributions corresponding to
 the Bonn (dot-dot-dashed),  Paris (dotted) and
Reid (dashed) potentials  are also shown.
}
\label{fig15}
\end{figure}

\begin{figure}   
\protect\caption{The  diagram defining the squared invariant matrix element
for the deuteron in terms of the
operator $\hat O$, eq. (\ref{ime0400}).
 The crossed line corresponds to
the nucleon on the mass-shell.}
\label{app1}
\end{figure}

\begin{figure} 
\caption{The  integration contour in eq. (\protect\ref{densit}).
The singularities labeled ``1'' and ``2'' 
correspond to the first and second  particles
on mass shell,  $p_{10}=w$ and $p_{20}=w$, respectively, 
whereas the singularities
labeled  ``3'' and ``4'' correspond to
  both particles off mass shell,
$p_{10}=-w$ and $p_{20}=-w$.}
\label{app2}
\end{figure}

\begin{figure} 
\caption{The deuteron densities: 
{\large a) }$\langle \gamma_0 \rangle (|{\bf  p}|) $.
The solid line corresponds to  the
integration over $p_0$, and the dashed line to the 
 nucleon pole contribution;
 the nucleon pole contribution to the scalar density
  $\langle 1 \rangle (|{\bf  p}|) $ is given by the 
  dotted line.
  {\large b) }$\langle 1 \rangle (|{\bf  p}|) $. 
  The solid line corresponds to the 
integration over $p_0$ and the dashed line to the 
 nucleon pole contribution;
 the dotted line represents
 the nucleon pole contribution to the scalar
 density calculated with 
 $\langle \gamma_0 \rangle (|{\bf  p}|)
 \left (1-{\bf  p}^2/(2m^2)\right ) $.}
\label{app3}
\end{figure}
\newpage


  \def\picsize{8cm}     
  \def\picfilename{pict1a.eps}
  \hspace*{-2cm}\epsfbox{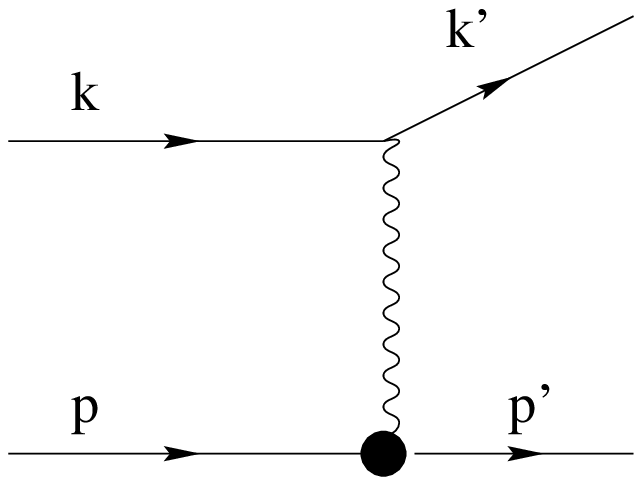}

\vskip -5.5cm


\def\picsize{6cm}
\def\picfilename{pict1b.eps}
 \epsfxsize\picsize
\hfill\epsfbox{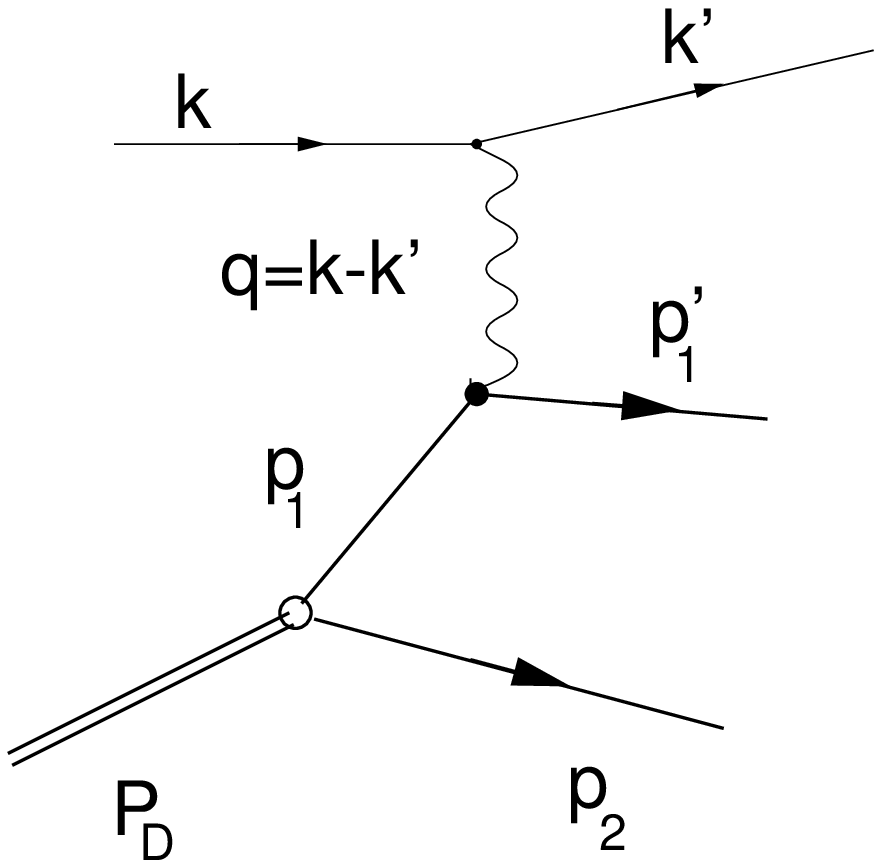}
 \vskip 2.2cm
\hspace{3cm} a) \hspace{10cm} b)
\vskip 1cm

Fig.~\ref{fig1}. 
C. Ciofi degli Atti.... Relativistic Structure of the Deuteron...

\vskip 1cm
\epsfxsize 6cm
\centerline{\epsfbox{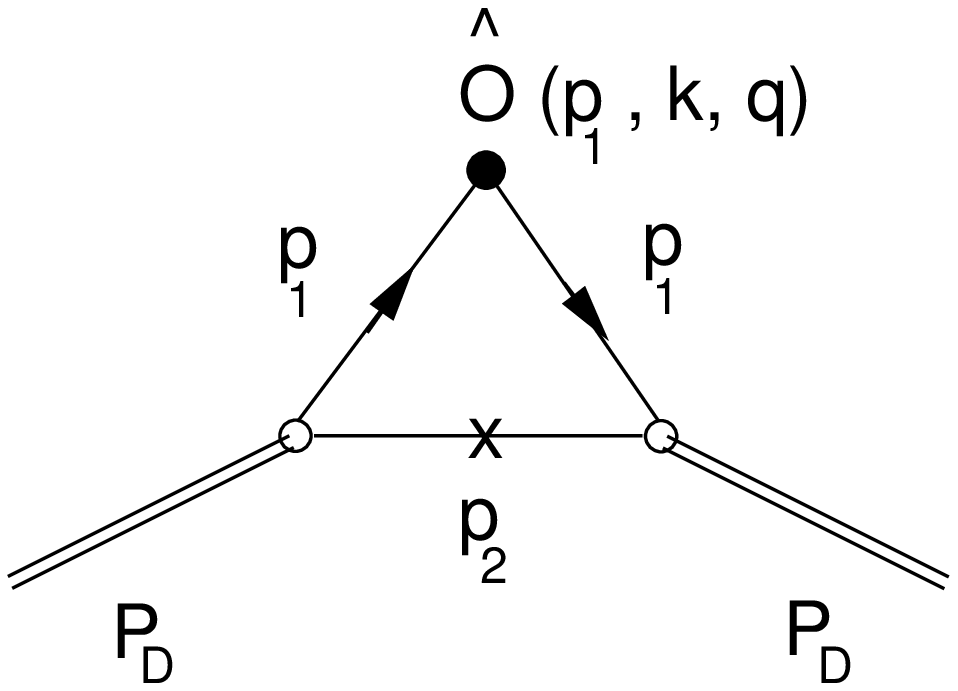}}

\vfill
Fig.~\ref{fig2}. 
C. Ciofi degli Atti.... Relativistic Structure of the Deuteron...

\vspace{-1cm}
\epsfxsize 14cm
\centerline{\epsfbox{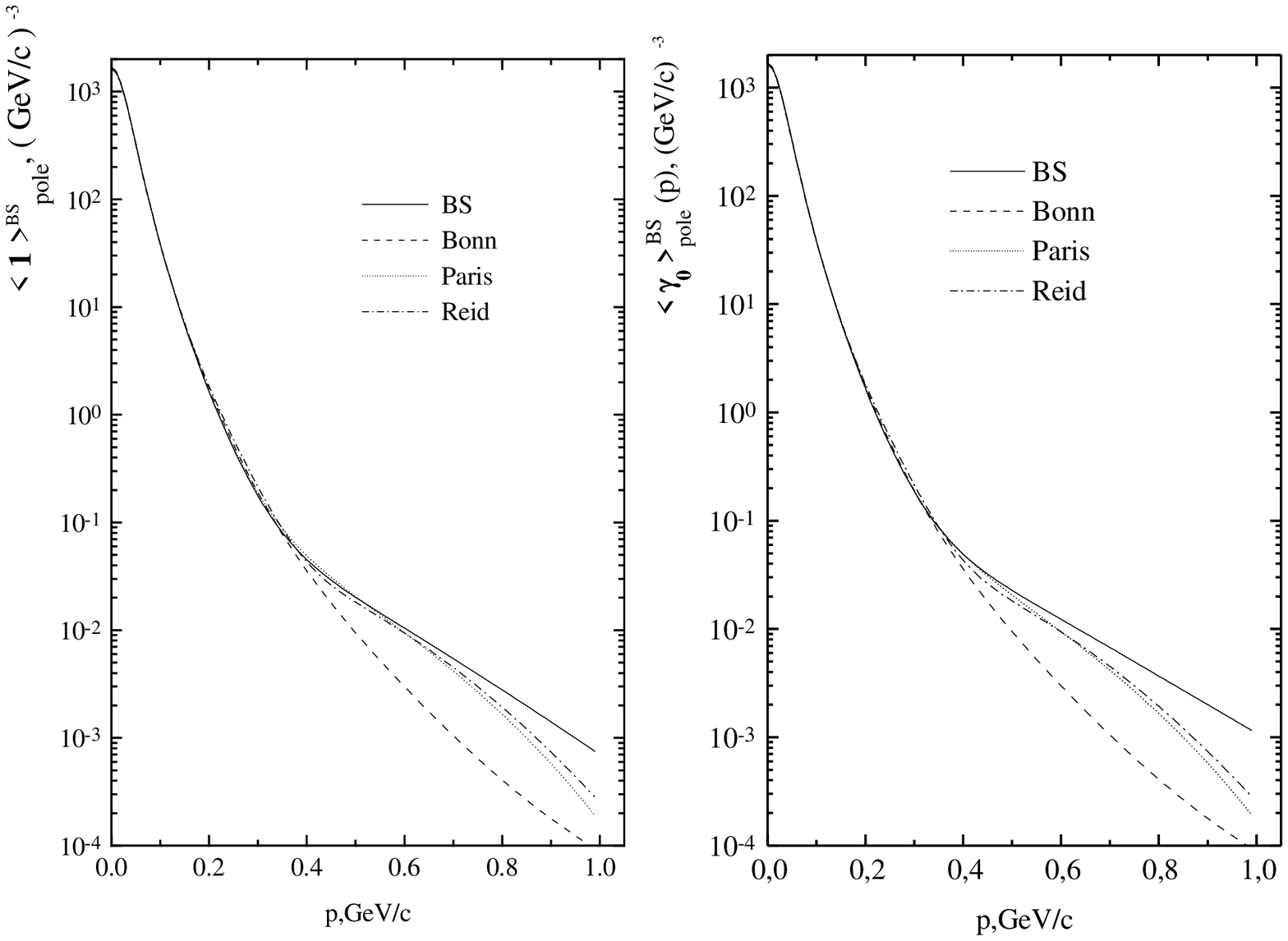}}

Fig.~\ref{fig3}. 
C. Ciofi degli Atti.... Relativistic Structure of the Deuteron...

 
\epsfxsize 8cm
\centerline{\epsfbox{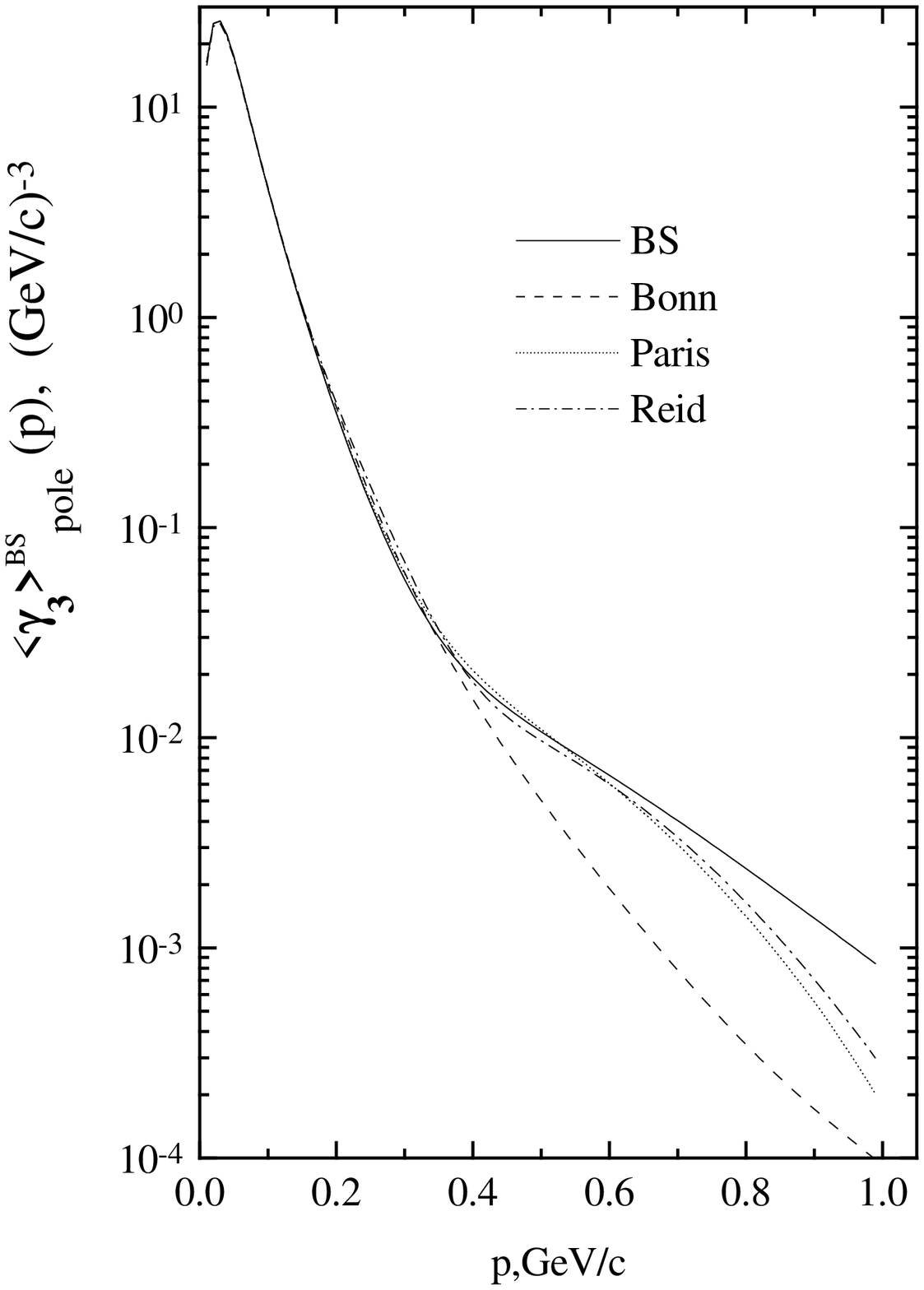}}

\vfill
Fig.~\ref{fig4}. 
C. Ciofi degli Atti.... Relativistic Structure of the Deuteron...

\newpage
 \vskip 5mm
\epsfxsize 16cm
\epsfbox{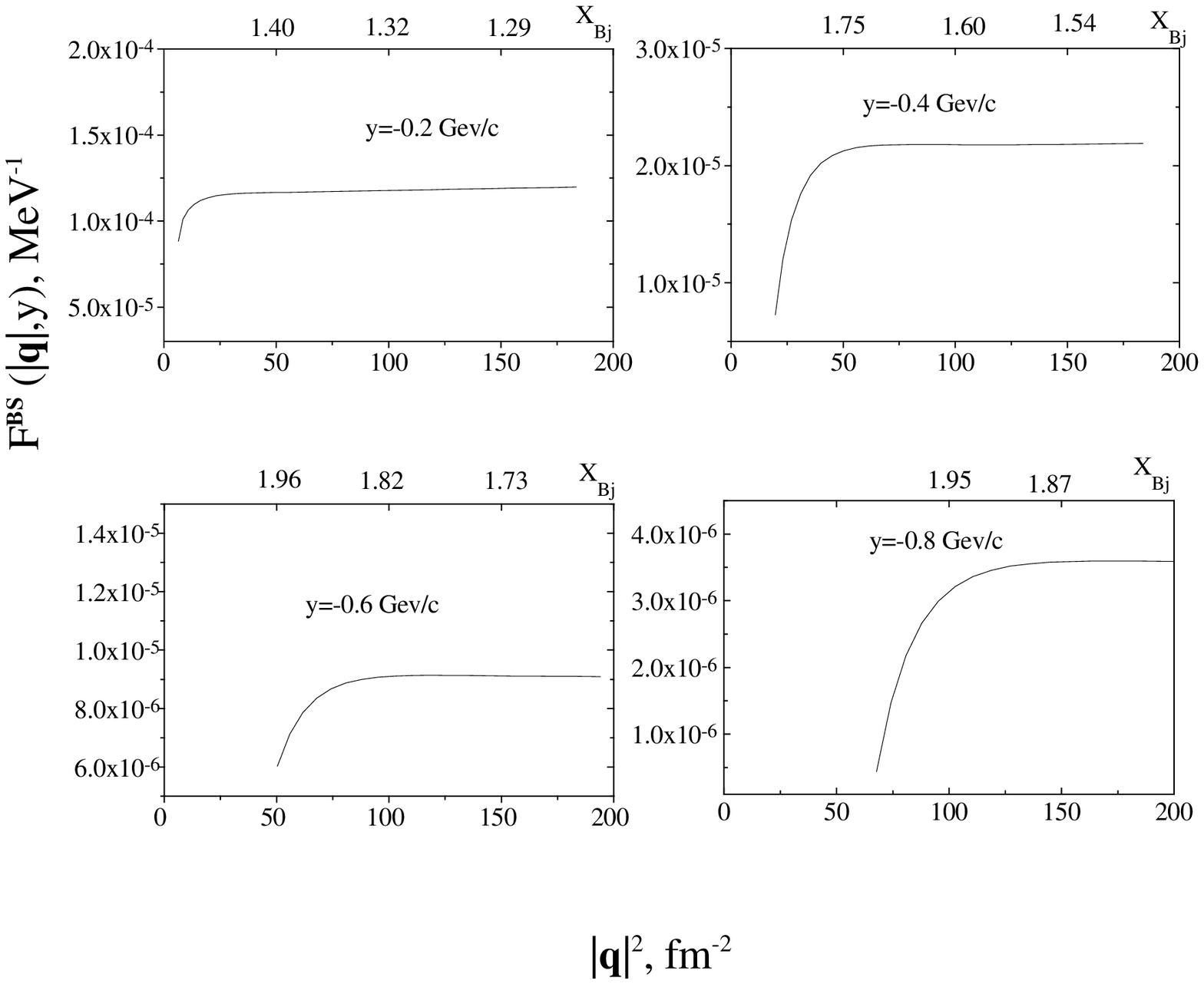}
 
\vfill
Fig.~\ref{fig5}. 
C. Ciofi degli Atti.... Relativistic Structure of the Deuteron...

\newpage
\vskip 5mm
\epsfxsize 12cm
\epsfbox{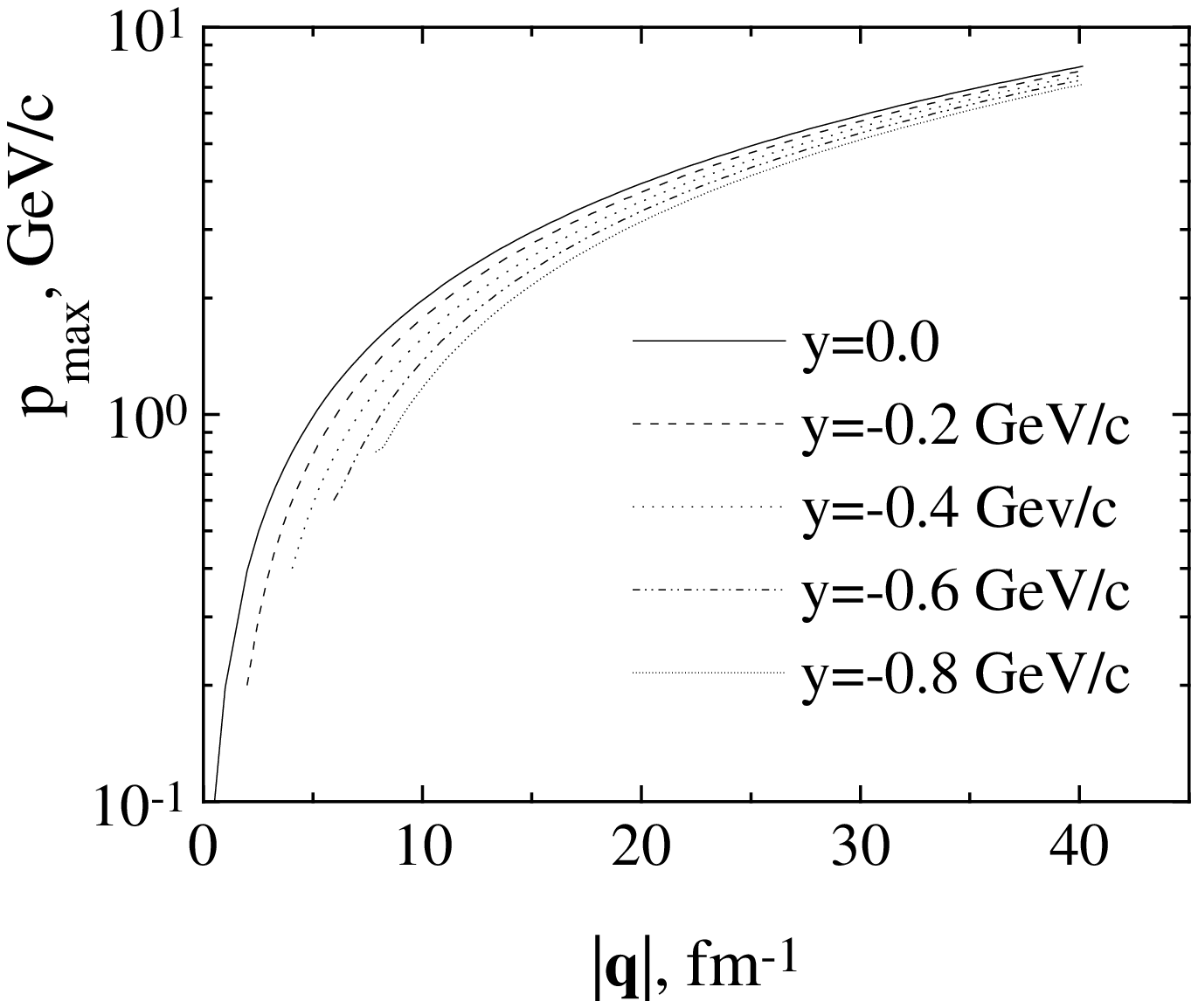}
  
\vfill
Fig.~\ref{fig6}. 
C. Ciofi degli Atti.... Relativistic Structure of the Deuteron...

\newpage
\epsfxsize 16cm
\epsfbox{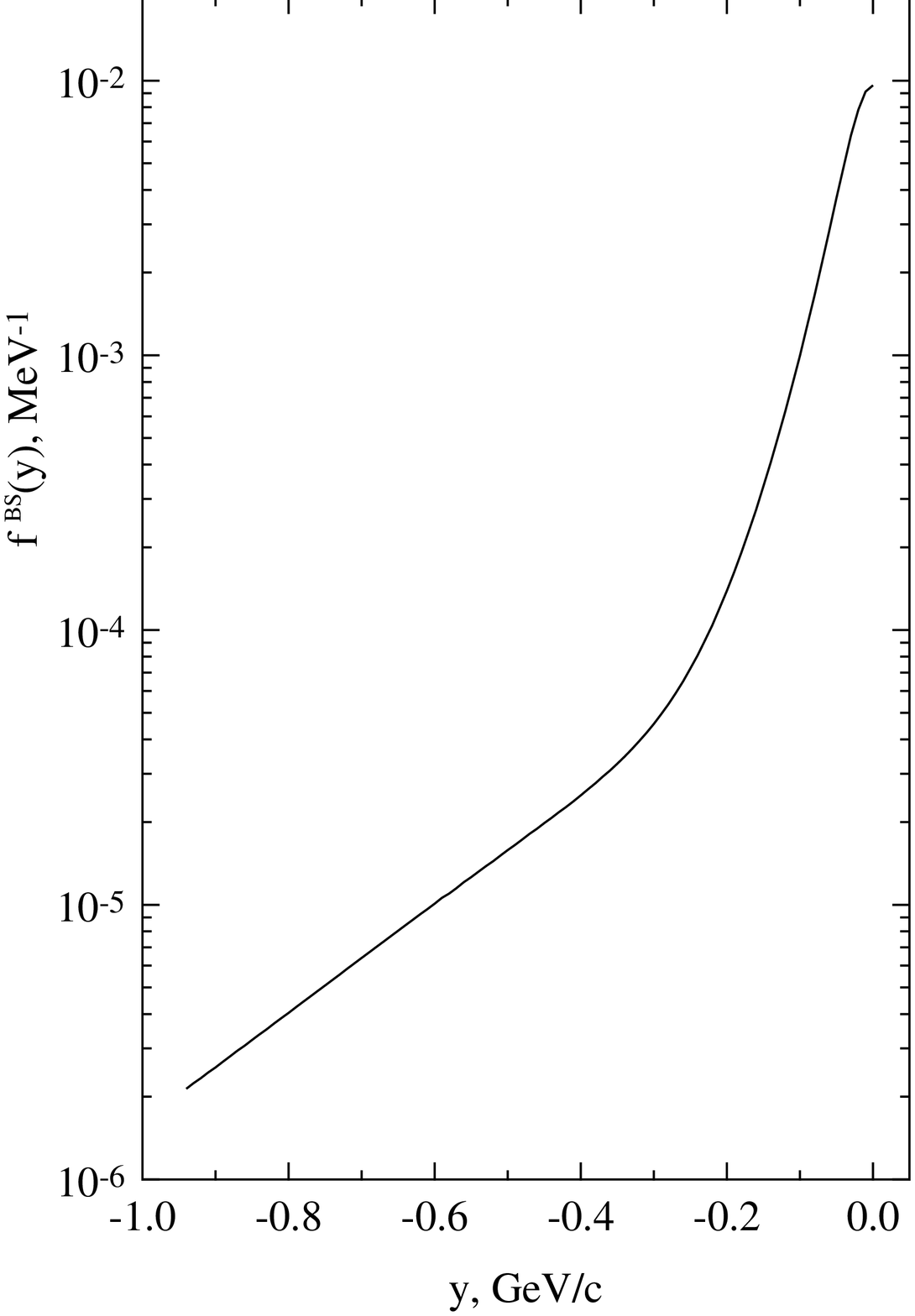}

\vfill
Fig.~\ref{fig7}. 
C. Ciofi degli Atti.... Relativistic Structure of the Deuteron...

\newpage
\epsfxsize 16cm
\epsfbox{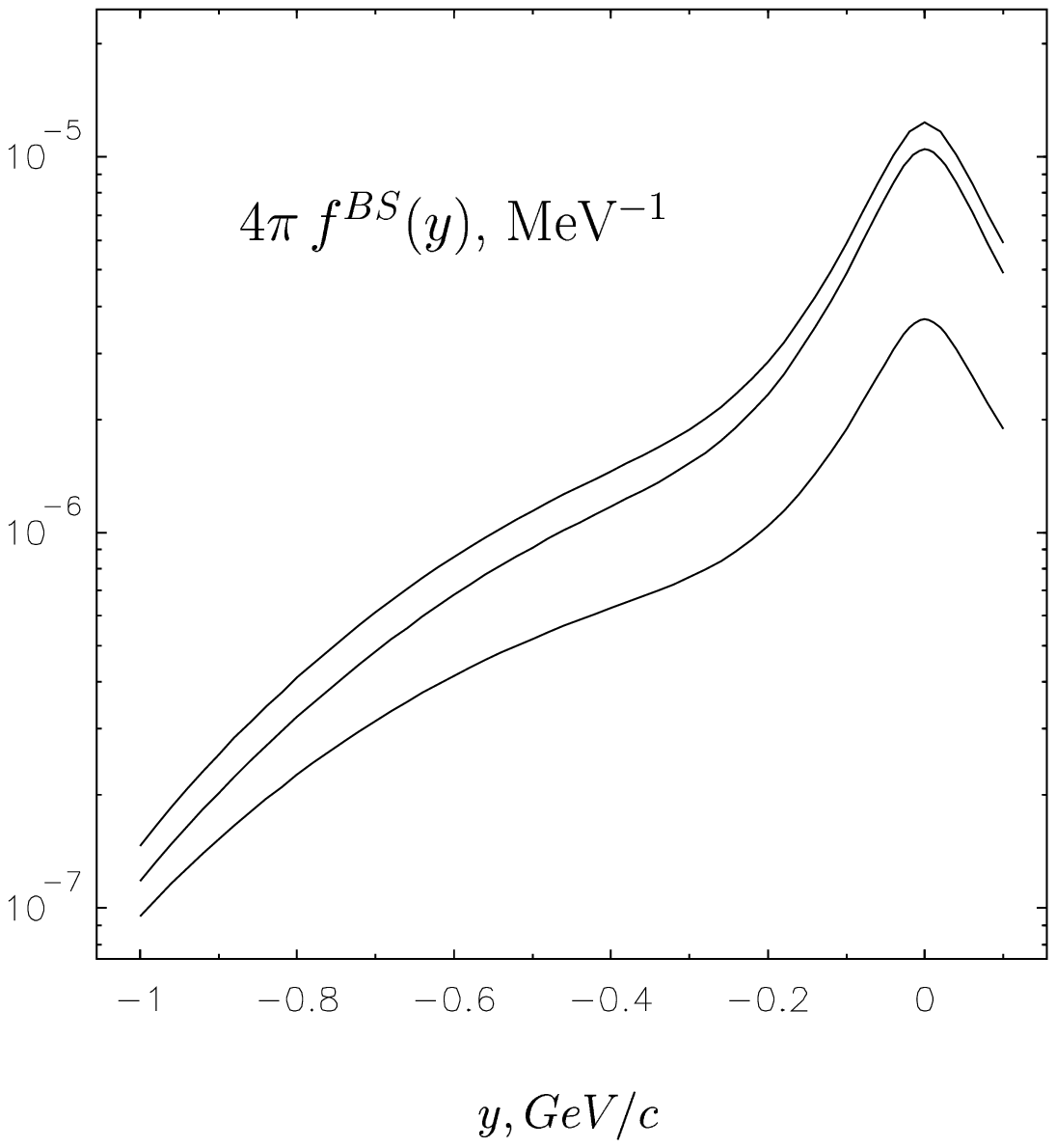}

Fig.~\ref{fig8}. 
C. Ciofi degli Atti.... Relativistic Structure of the Deuteron...

\epsfxsize 12cm
\epsfbox{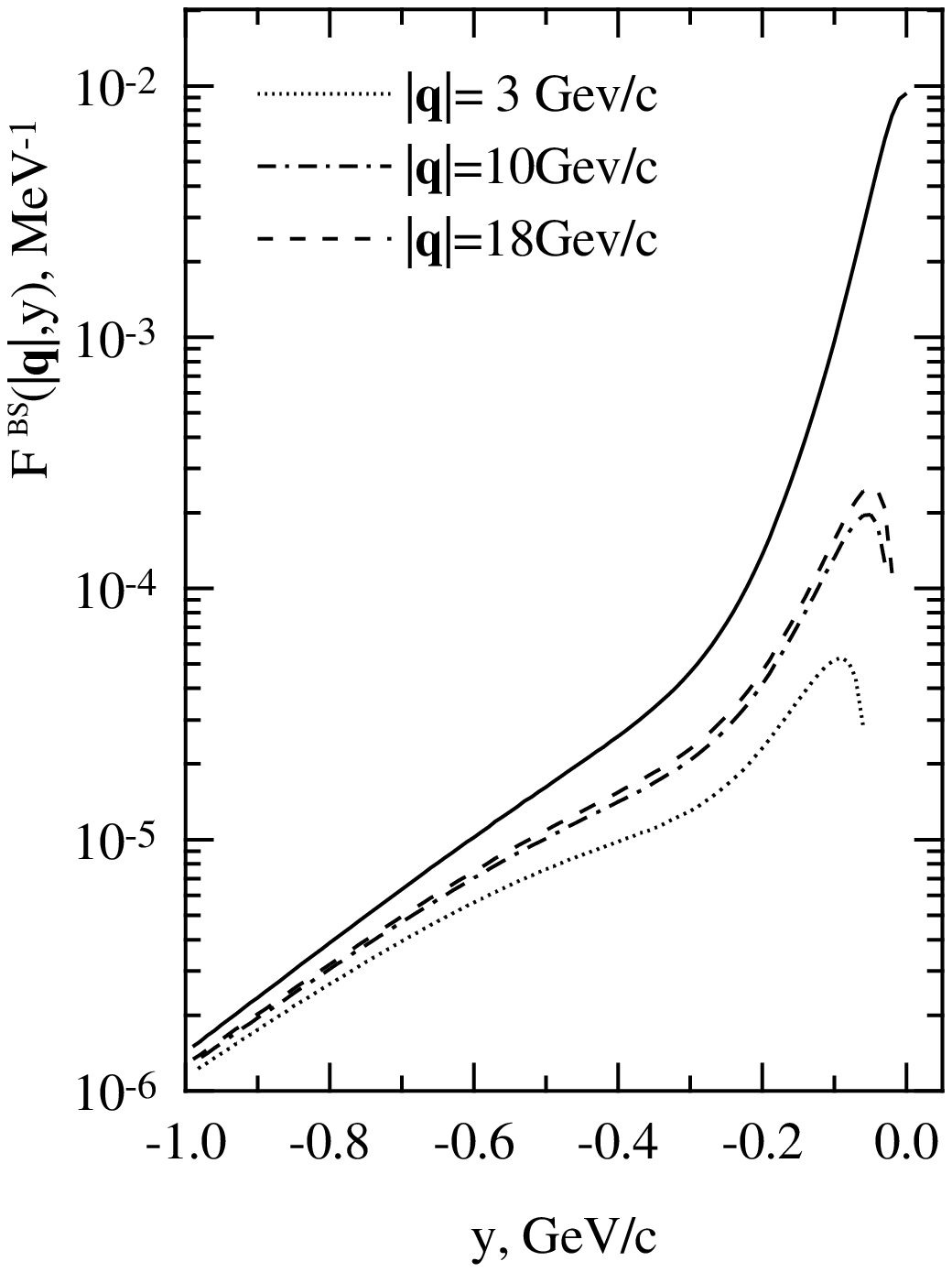}
\vskip 2cm

Fig.~\ref{fig9}. 
C. Ciofi degli Atti.... Relativistic Structure of the Deuteron...

\newpage

\epsfxsize 16cm
\epsfbox{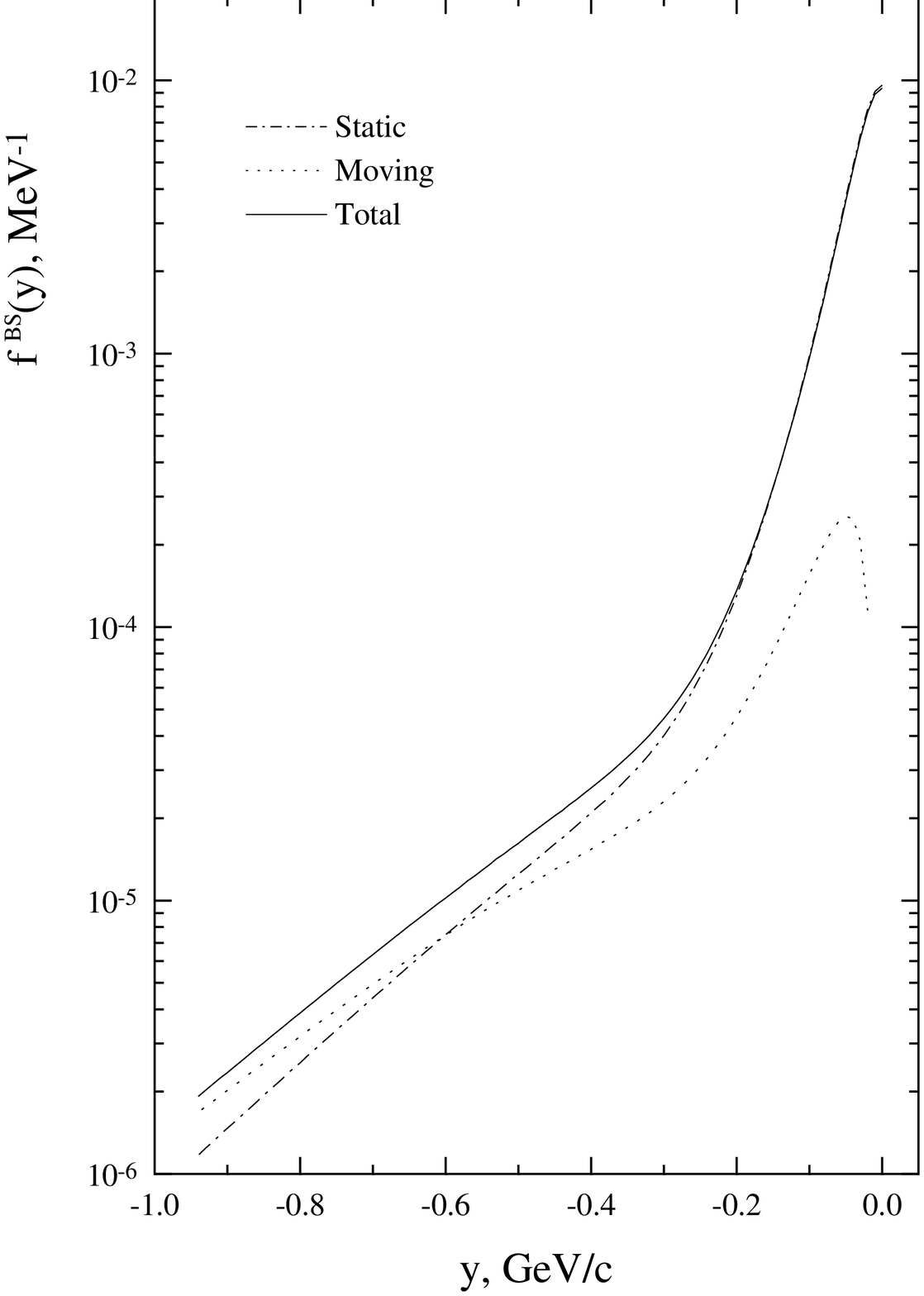}
  
\vfill
Fig.~\ref{fig10}. 
C. Ciofi degli Atti.... Relativistic Structure of the Deuteron...

\newpage
\epsfxsize 16cm
\epsfbox{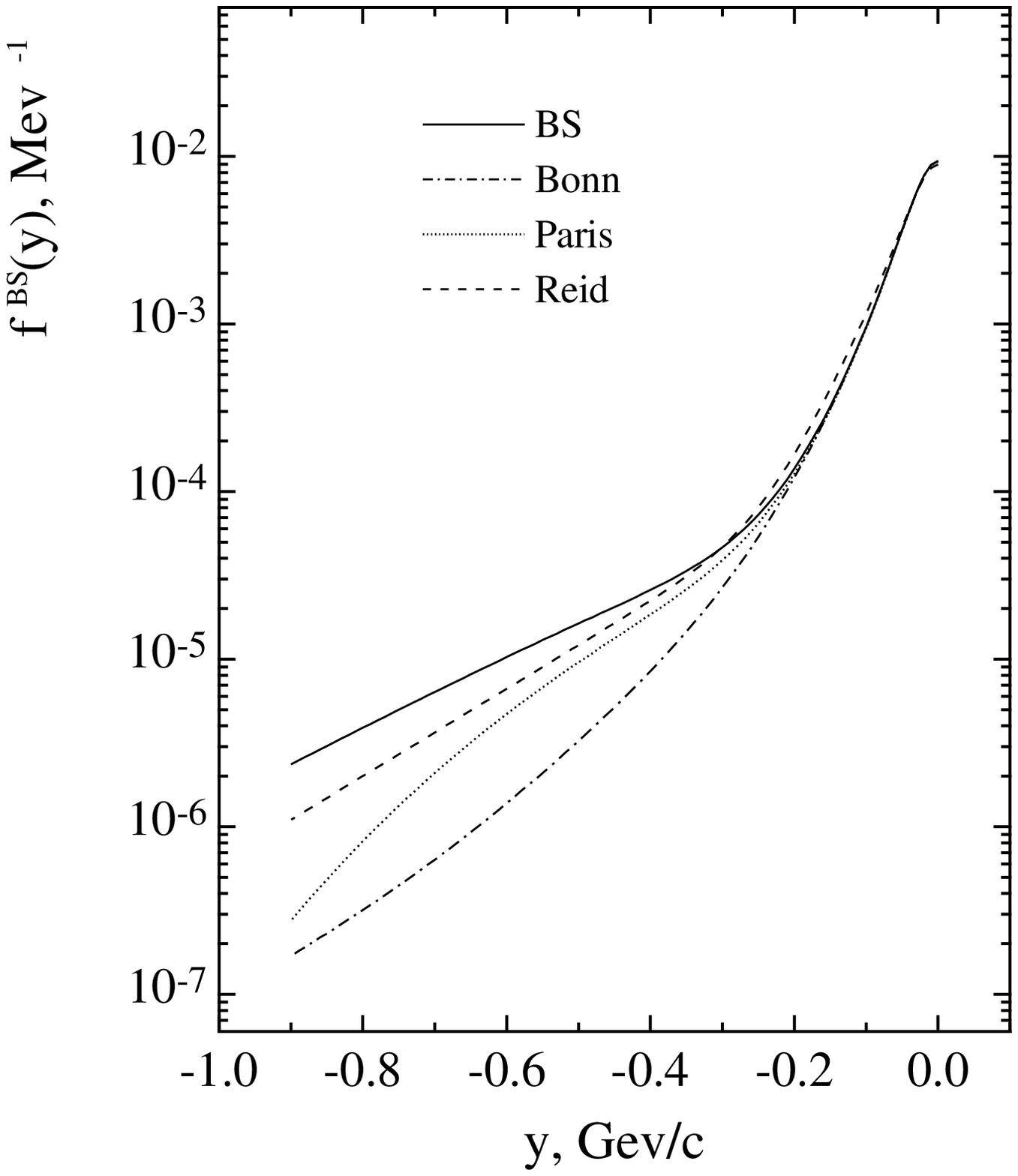}

\vfill
Fig.~\ref{fig11}. 
C. Ciofi degli Atti.... Relativistic Structure of the Deuteron...

\newpage
\epsfxsize 14cm
\epsfbox{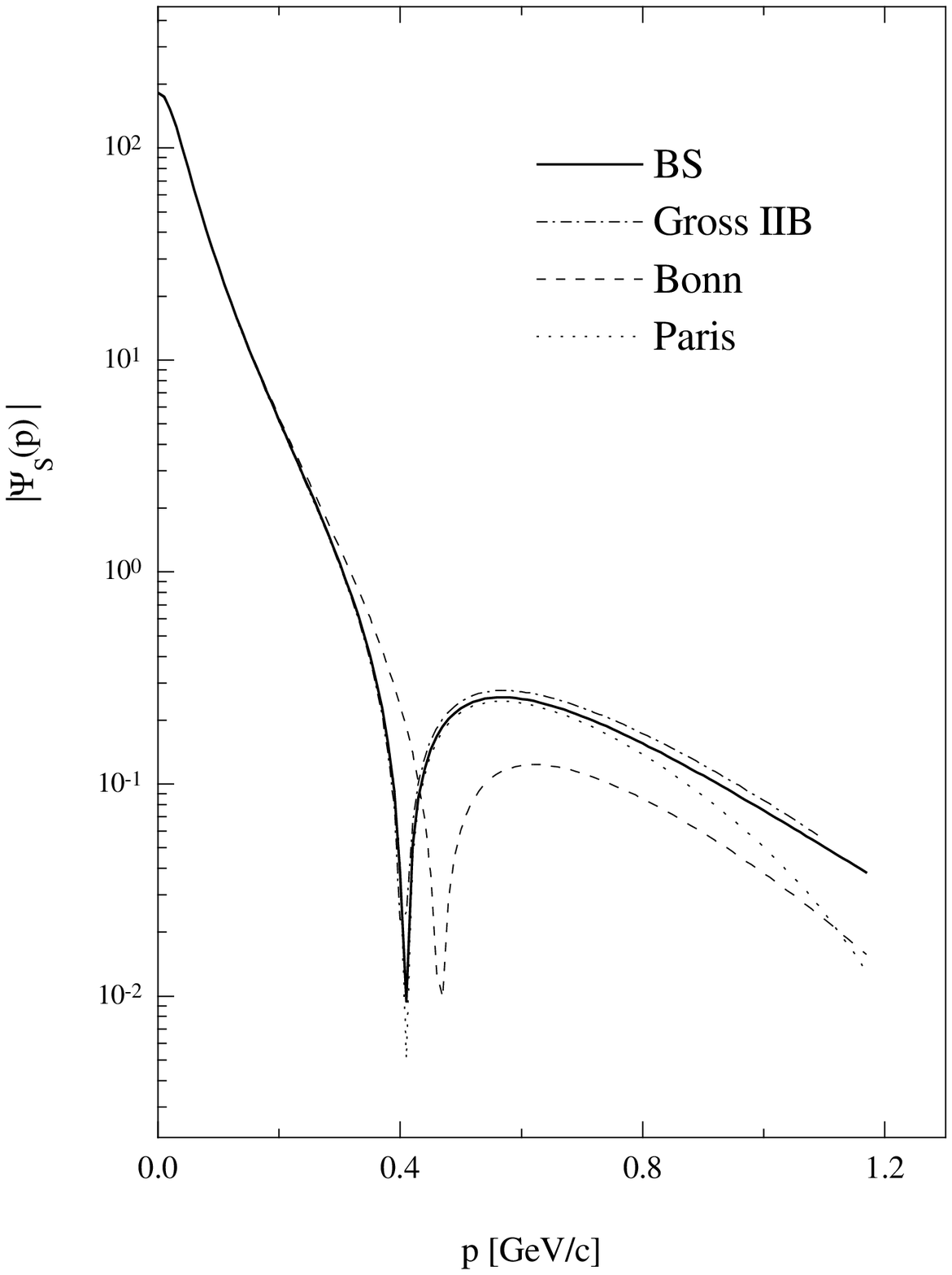}

\vfill
Fig.~\ref{fig12}. 
C. Ciofi degli Atti.... Relativistic Structure of the Deuteron...

\newpage
\epsfxsize 14cm
\epsfbox{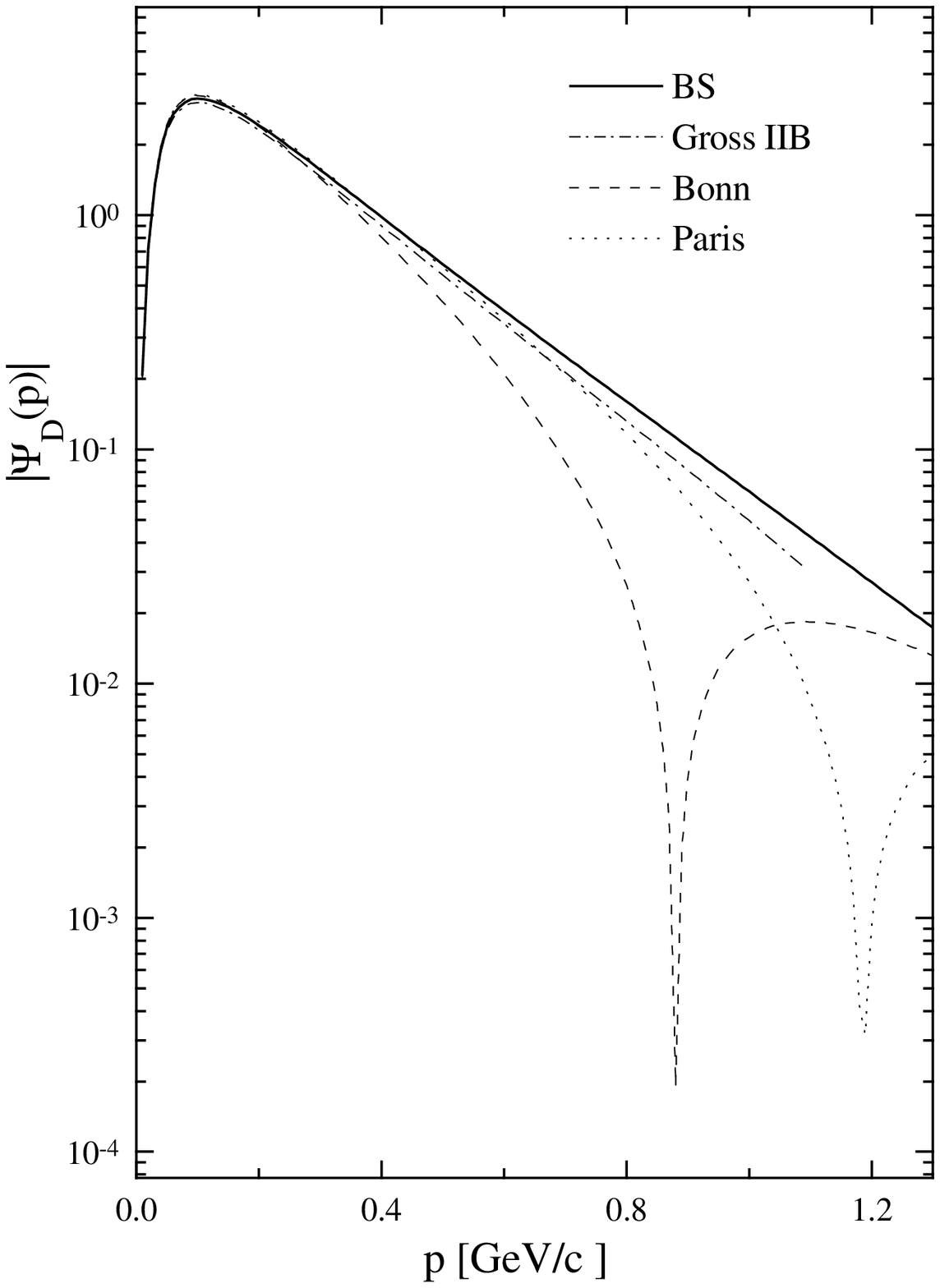}

\vfill
Fig.~\ref{fig13}. 
C. Ciofi degli Atti.... Relativistic Structure of the Deuteron...

\newpage

\epsfxsize 16cm
\epsfbox{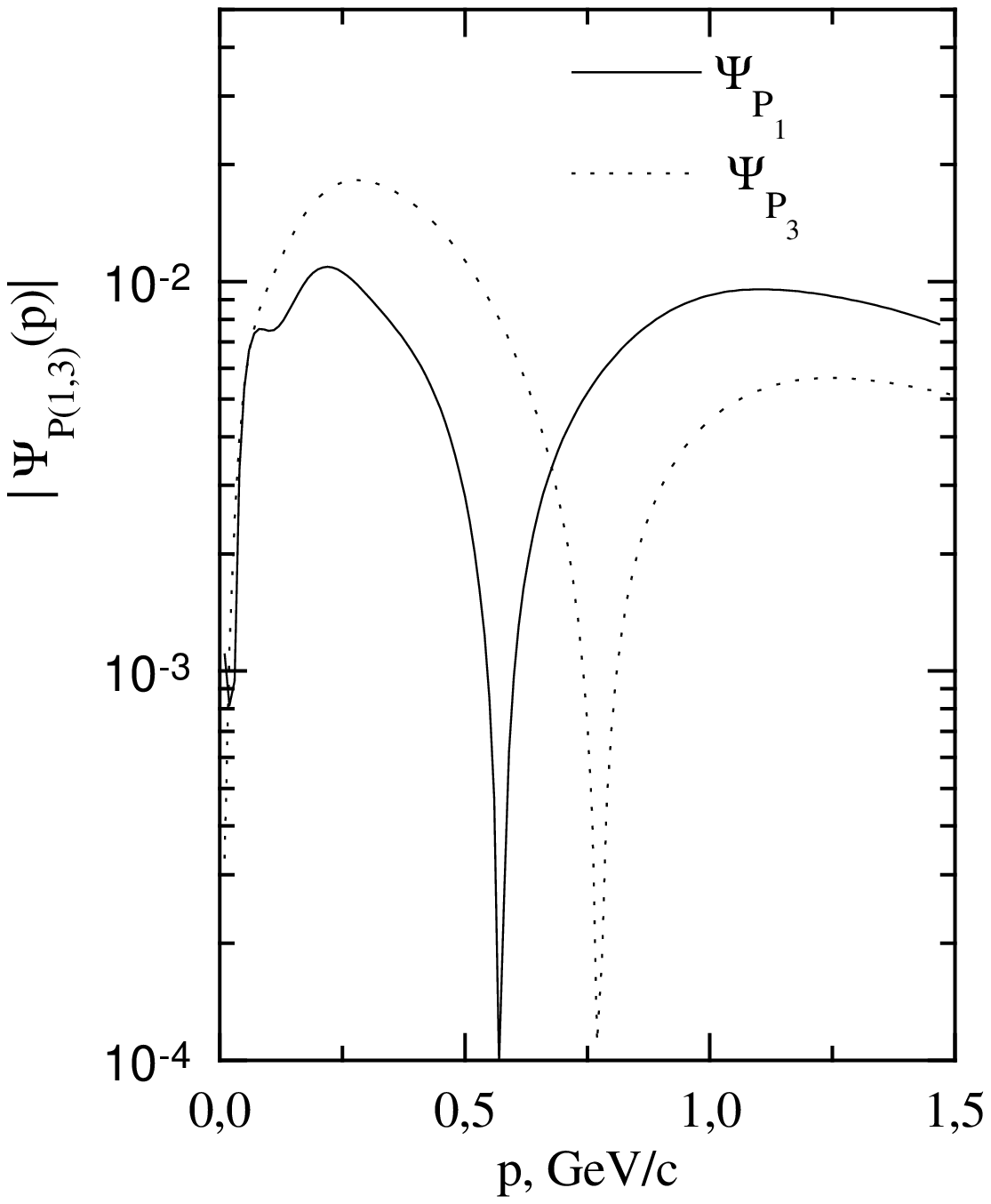}
 
\vfill
Fig.~\ref{fig14}. 
C. Ciofi degli Atti.... Relativistic Structure of the Deuteron...

\newpage
\epsfxsize 12cm
\epsfbox{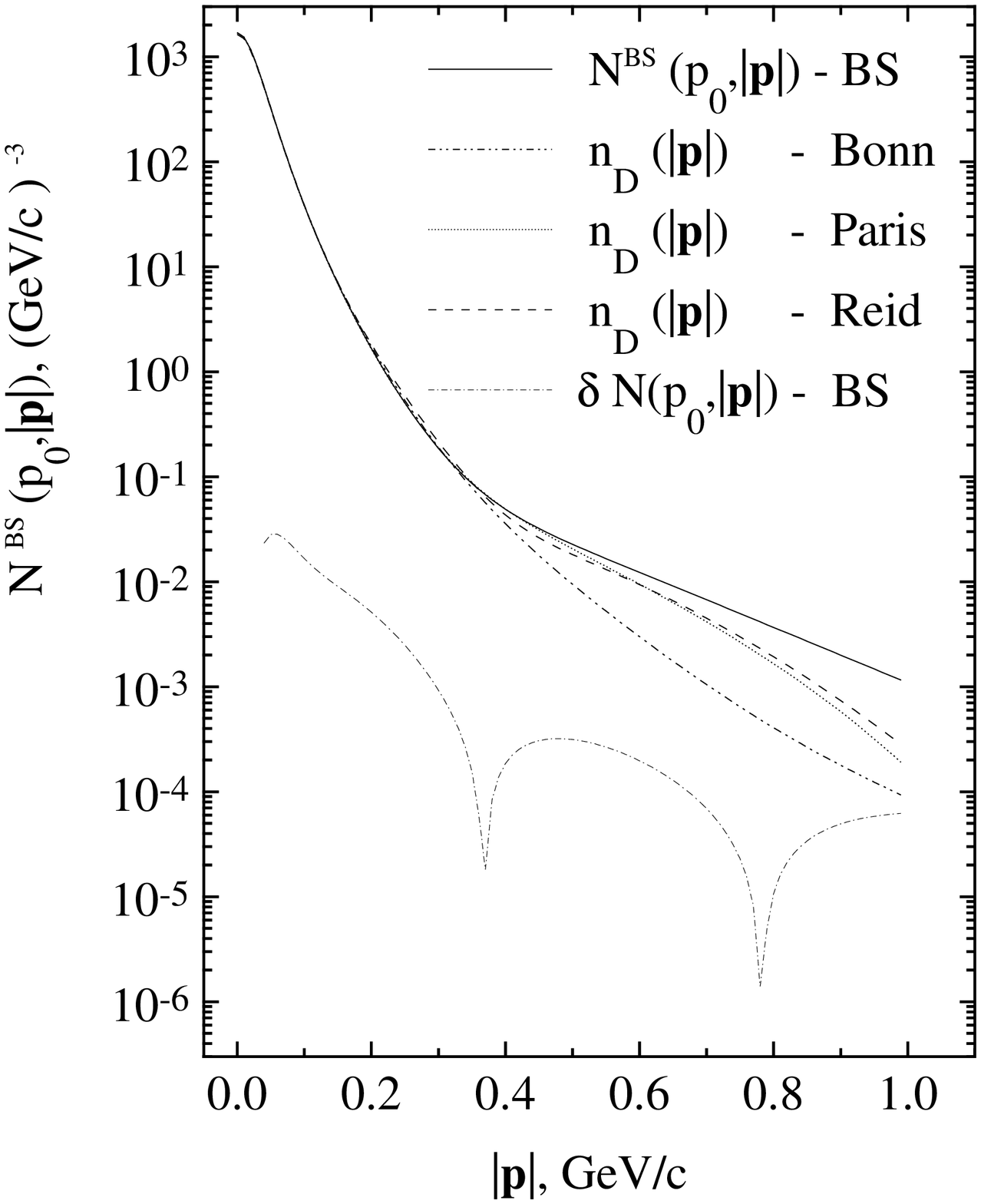}

\vfill
Fig.~\ref{fig15}. 
C. Ciofi degli Atti.... Relativistic Structure of the Deuteron...

\newpage
\epsfxsize 12cm
\epsfbox{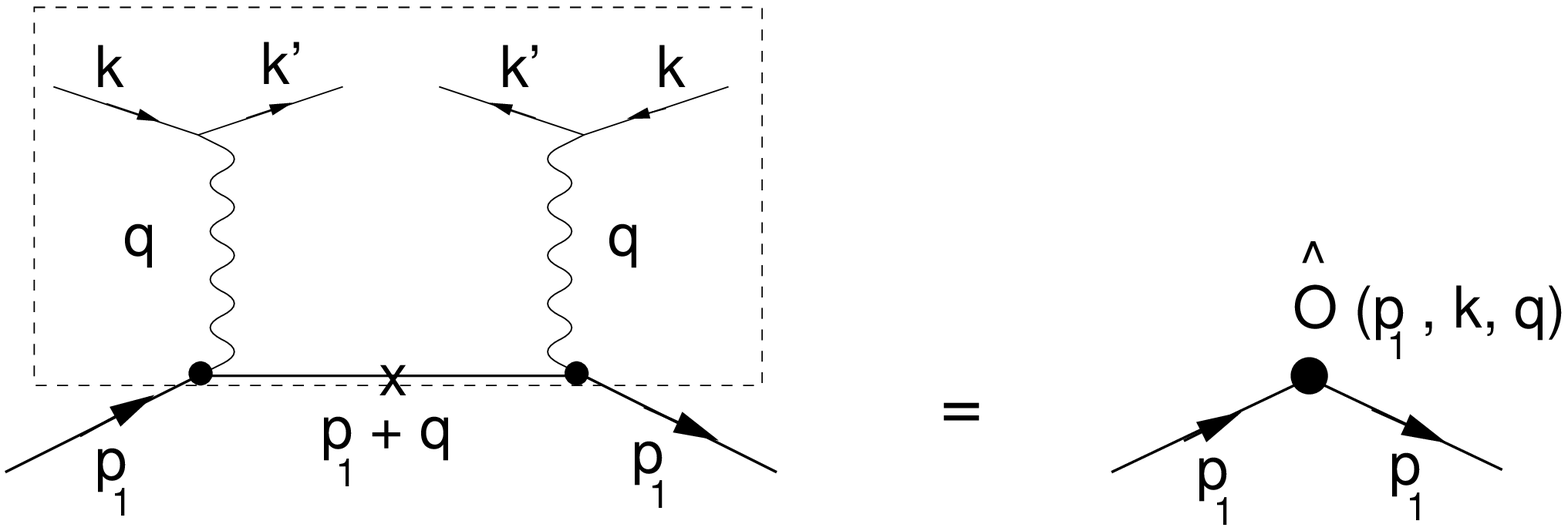}

\vfill
Fig.~\ref{app1}. 
C. Ciofi degli Atti.... Relativistic Structure of the Deuteron...

\newpage
\epsfxsize 12cm
\epsfbox{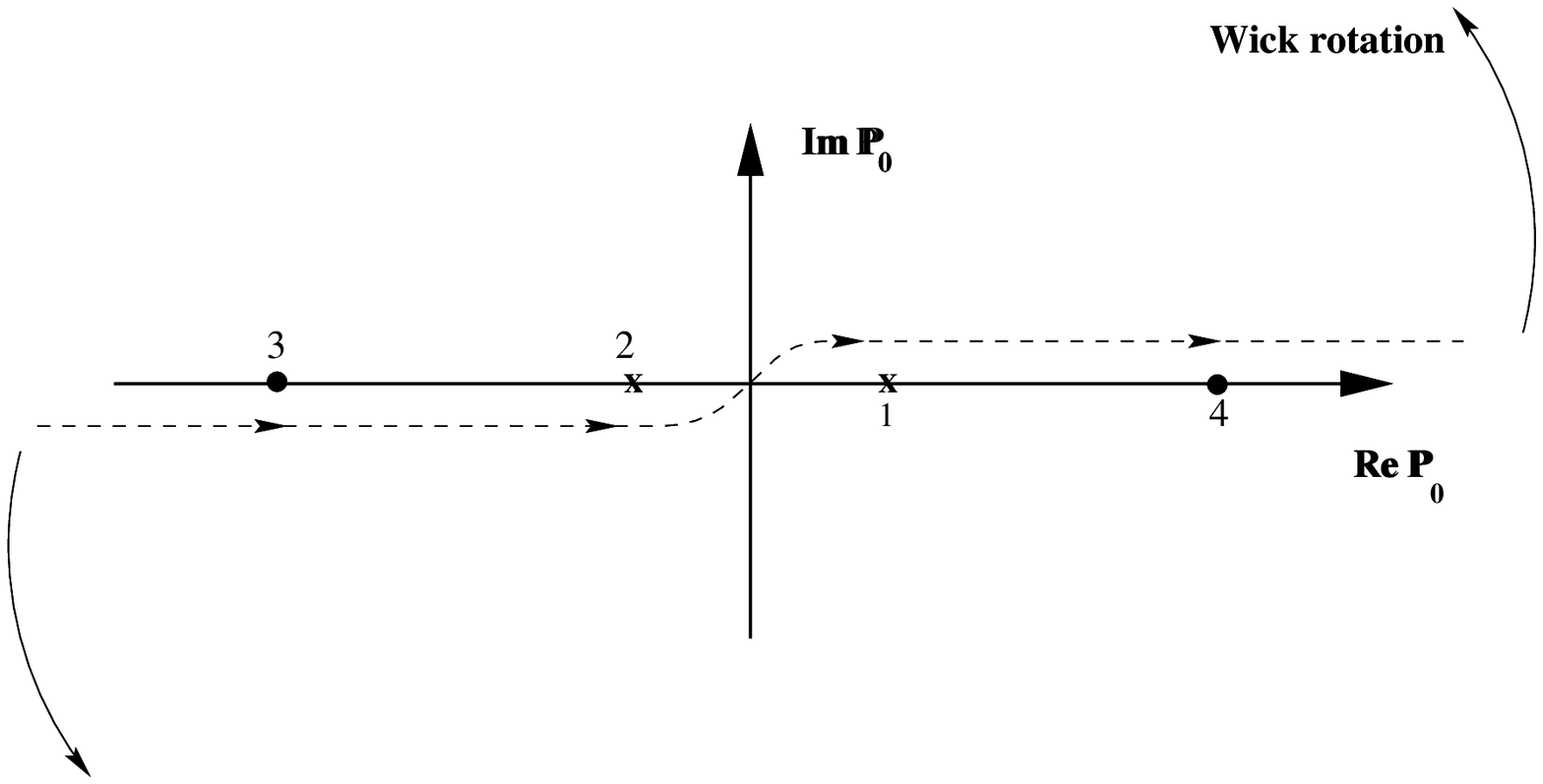}

\vfill
Fig.~\ref{app2}. 
C. Ciofi degli Atti.... Relativistic Structure of the Deuteron...

\newpage
\epsfxsize 12cm
\epsfbox{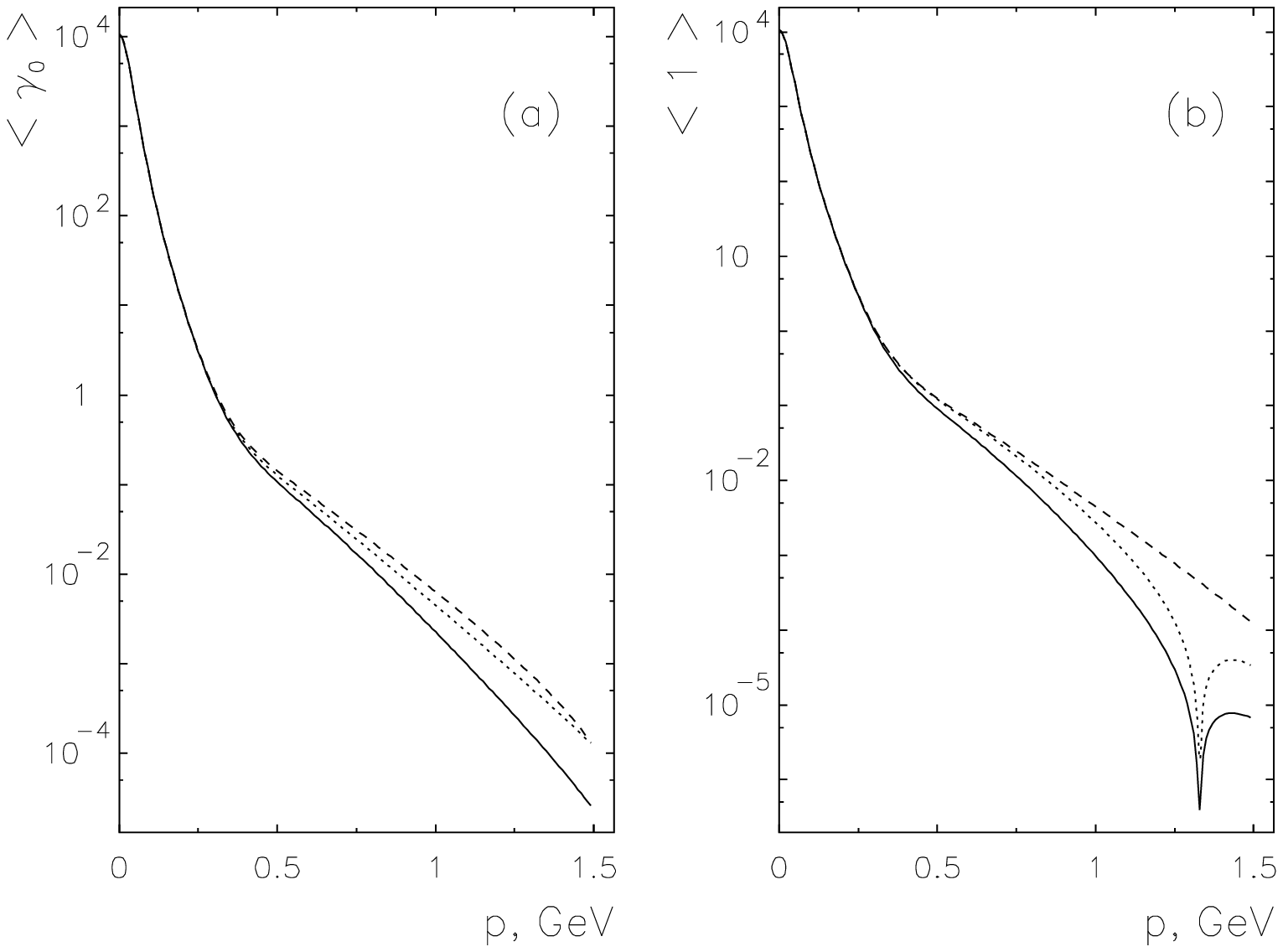}
 
\vfill
Fig.~\ref{app3}. 
C. Ciofi degli Atti.... Relativistic Structure of the Deuteron...
 \end{document}